\documentclass[aps,prb,twocolumn,amsmath,amssymb,showpacs,superscriptaddress]{revtex4-1}

\usepackage{graphicx}
\usepackage{amsmath,amssymb,amsfonts}
\usepackage{textcomp}
\usepackage{hyperref}
\usepackage{gensymb}
\usepackage{verbatim}
\usepackage{amsmath}
\hypersetup{breaklinks=true,colorlinks=true,urlcolor=black}
\usepackage{color}
\usepackage{soul}
\DeclareGraphicsExtensions{.jpg,.pdf,.png,.eps}

\begin{document}
\title{Pressure-temperature phase diagrams of CaK(Fe$_{1-x}$Ni$_{x}$)$_4$As$_4$ superconductors}
\author{Li Xiang}
\affiliation{Ames Laboratory, Iowa State University, Ames, Iowa 50011, USA}
\affiliation{Department of Physics and Astronomy, Iowa State University, Ames, Iowa 50011, USA}
\email[]{ives@iastate.edu}
\author{William R. Meier}
\affiliation{Ames Laboratory, Iowa State University, Ames, Iowa 50011, USA}
\affiliation{Department of Physics and Astronomy, Iowa State University, Ames, Iowa 50011, USA}
\author{Mingyu Xu}
\affiliation{Ames Laboratory, Iowa State University, Ames, Iowa 50011, USA}
\affiliation{Department of Physics and Astronomy, Iowa State University, Ames, Iowa 50011, USA}
\author{Udhara S. Kaluarachchi}
\affiliation{Ames Laboratory, Iowa State University, Ames, Iowa 50011, USA}
\affiliation{Department of Physics and Astronomy, Iowa State University, Ames, Iowa 50011, USA}
\author{Sergey L. Bud'ko}
\affiliation{Ames Laboratory, Iowa State University, Ames, Iowa 50011, USA}
\affiliation{Department of Physics and Astronomy, Iowa State University, Ames, Iowa 50011, USA}
\author{Paul C. Canfield}
\affiliation{Ames Laboratory, Iowa State University, Ames, Iowa 50011, USA}
\affiliation{Department of Physics and Astronomy, Iowa State University, Ames, Iowa 50011, USA}
\email[]{canfield@ameslab.gov}

\begin{abstract}
	The pressure dependence of the magnetic and superconducting transitions, as well as that of the superconducting upper critical field is reported for CaK(Fe$_{1-x}$Ni$_{x}$)$_4$As$_4$, the first example of an Fe-based superconductor with spin-vortex-crystal-type magnetic ordering. Resistance measurements were performed on single crystals with two substitution levels ($x=0.033, 0.050$) under hydrostatic pressures up to 5.12 GPa and in magnetic fields up to 9 T. Our results show that, for both compositions, magnetic transition temperatures, $T_\textrm{N}$, are suppressed upon applying pressure, the superconducting transition temperatures $T_\textrm{c}$ are suppressed by pressure as well, except for $x=0.050$ in the pressure region where $T_\textrm{N}$ and $T_\textrm{c}$ cross. Furthermore, the pressure associated with the crossing of the $T_\textrm{N}$ and $T_\textrm{c}$ lines also coincides with a minimum in the normalized slope of the superconducting upper critical field, consistent with a likely Fermi-surface reconstruction associated with the loss of magnetic ordering. Finally, at $p \sim$ 4 GPa, both Ni-substituted CaK(Fe$_{1-x}$Ni$_{x}$)$_4$As$_4$ samples likely go through a half-collapsed-tetragonal (hcT) phase transition, similar to the parent compound CaKFe$_4$As$_4$.
	
\end{abstract}

\maketitle 

\section{Introduction}
Since the discovery of Fe-based superconductors (FeSC) \cite{Kamihara2008,Ren2008,Rotter2008PRL,Takahashi2008Nat}, many studies have been done on them and they have expanded into a large family. Among them the $Ae$Fe$_2$As$_2$ compounds ($Ae$=Ca, Sr, Ba, Eu) have received significant attention because large, high-quality single crystals can be obtained with a variety of chemical substitution\cite{Canfield2010,Ni2011}. Studies have revealed that members of this family share a global phase diagram upon tuning by substitution or pressure\cite{Paglione2010,Canfield2010}. At ambient pressure, the parent compounds undergo a structural/magnetic transition upon cooling; substitution or pressure induce superconductivity after sufficiently suppressing the structural/magnetic transitions\cite{Canfield2010,Ni2011,Torikachvili2008PRB,Alireza2009,Kimber2009,Colombier2009}. This suggests a competition between the magnetism and superconductivity, and that magnetic fluctuations play an important role in forming superconductivity in this system\cite{Pratt2009PRL,Christianson2009PRL,Fernandes2010PRB,Christianson2008Nat,Yu2009Nat,Paglione2010}.

Recently, a new FeSC $AeA$Fe$_4$As$_4$ ($A$=K, Rb, Cs) structural type ($P4/mmm$) was discovered by Iyo $et$ $al$\cite{Iyo2016}. This is not a homogeneous substitution as in ($Ae_{0.5}A_{0.5}$)Fe$_2$As$_2$ where $Ae/A$ share the same crystallographic site. Each $Ae$ and $A$ in the $AeA$Fe$_4$As$_4$ structure has a unique, well-defined, crystallographic site, forming alternating $Ae$ and $A$ planes along the $c$-axis\cite{Iyo2016, Meier2016PRB}. Among them, single crystals of CaKFe$_4$As$_4$ were synthesized and found to be superconducting at $\sim$ 35 K and no other phase transition from 1.8 K to 300 K at ambient pressure\cite{Meier2016PRB,Meier2017PRM}. A pressure study up to 6 GPa shows that the superconducting transition temperature, $T_\textrm{c}$, is suppressed to about 28.5 K before it undergoes half-collapsed-tetragonal (hcT) phase transition at $\sim$ 4 GPa and loses bulk superconductivity\cite{Kaluarachchi2017PRB}. The hcT phase transition occurs due to the As-As bonding across the Ca-layer under pressure, like the collapsed-tetragonal transition in CaFe$_2$As$_2$ at $\sim$0.35 GPa\cite{Torikachvili2008PRL,Yu2009PRB,Kreyssig2008PRB}.

From the perspective of electron count, CaKFe$_4$As$_4$ is analogous to (Ba$_{0.5}$K$_{0.5}$)Fe$_2$As$_2$ and many of its properties are consistent with this\cite{Meier2016PRB}. In the later compound, the stripe-type spin density wave associated with BaFe$_2$As$_2$ is suppressed by hole doping\cite{Paglione2010} (substitution K for Ba). A recent study revealed that adding electrons to CaKFe$_4$As$_4$ via Ni or Co substitution drives the system back towards a magnetic phase. In contrast to the stripe-type antiferromagnetism in the "122" systems, the order in the Ni- or Co-substituted CaKFe$_4$As$_4$ is experimentally identified as a new hedgehog spin-vortex-crystal (SVC) magnetism that has no structural phase transition associated with it\cite{Meier2018}. This type of magnetic order had been theoretically predicted but until the discovery of Ni- or Co-substituted CaKFe$_4$As$_4$, was considered to be a "missing link"\cite{Fernandes2016PRB,Cvetkovic2013PRB,Halloran2017PRB}. Increasing the substitution level of Ni or Co in CaK(Fe$_{1-x}T_x$)$_4$As$_4$ leads to the suppression of the superconducting transition temperature $T_\textrm{c}$ and stabilizing the SVC magnetism and increasing $T_\textrm{N}$\cite{Meier2018}.

The application of pressure to Ba(Fe$_{1-x}$Co$_x$)$_2$As$_2$ suppresses AFM ($T_\textrm{N}$ falls) and increases $T_\textrm{c}$\cite{Colombier2010}. This has been taken as an indication that pressure, like doping, can tune $T_\textrm{N}$ and the associated AFM fluctuations to favor the superconducting state when $T_\textrm{N} >T_\textrm{c}$. Therefore, it is natural to study how the SVC magnetic order behaves under pressure, specifically, how the magnetism and superconductivity interact in this system and whether this interaction is similar to Ba(Fe$_{1-x}$Co$_x$)$_2$As$_2$.

In this work, we present the first pressure study on Ni-substituted CaK(Fe$_{1-x}$Ni$_{x}$)$_4$As$_4$ ($x=0.033$ and $0.050$) up to 5.12 GPa. The pressure-temperature ($p-T$) phase diagrams inferred from resistance measurements allow comparison of $T_\textrm{N}(p)$ and $T_\textrm{c}(p)$. Specifically, $p-T$ phase diagrams reveal that $T_\textrm{N}$ is suppressed with pressure for both substitution levels. In contrast to Ba(Fe$_{1-x}$Co$_x$)$_2$As$_2$, $T_\textrm{c}$ is suppressed as well, although more slowly. For $x=0.050$, it exhibits an anomaly at the pressure where $T_\textrm{c}$ and $T_\textrm{N}$ cross. At $\sim$ 4 GPa both compositions appear to undergo the hcT transition as was observed in the undoped CaKFe$_4$As$_4$. Furthermore, superconducting upper critical fields studied up to 9 T suggests a Fermi-surface reconstruction when $T_\textrm{N}(p)$ crosses $T_\textrm{c}(p)$.

\section{Experimental details}
Single crystals of CaK(Fe$_{1-x}$Ni$_{x}$)$_4$As$_4$ ($x=0.033$ and $0.050$) with sharp superconducting transitions at ambient pressure [See Figs 1(b)-3(b)] were grown using high-temperature solution growth\cite{Meier2016PRB,Meier2017PRM}. The substitution level, $x$, was determined by performing wavelength-dispersive x-ray spectroscopy (WDS) as described in Ref. \onlinecite{Meier2018}.

The in-plane $ab$ resistance was measured using standard four-probe configuration. The 25 $\mu$m Pt wires were soldered to the samples using a Sn:Pb-60:40 alloy. For $x=0.033$, two samples, $\#$1 and $\#$2, were cut from one single crystal. They were then measured in a piston-cylinder cell (PCC)\cite{Budko1984} and a modified Bridgman Anvil Cell (mBAC)\cite{Colombier2007} respectively. For $x=0.050$, a single sample was prepared and measured in the mBAC. Pressure values for both cells, at low temperature, were inferred from the $T_\textrm{c}(p)$ of lead\cite{Bireckoven1988}. For the PCC, a 4:6 mixture of light mineral oil:n-pentane was used as the pressure medium, which solidifies, at room temperature, in the range of 3-4 GPa. For the mBAC, a 1:1 mixture of iso-pentane:n-pentane was used as the pressure medium, which solidifies, at room temperature, in the range of 6-7 GPa. Both of the solidification pressures are well above the maximum pressures achieved in the pressure cells, which suggests good hydrostatic conditions\cite{Budko1984,Kim2011,Torikachvili2015}.

The ac resistance measurements were performed in a Quantum Design Physical Property Measurement System using $I=1$ mA; $f=17$ Hz excitation, on cooling with the rate of 0.25 K/min and the magnetic field was applied along the $c$ axis.

\section{Results and discussions}
	
\begin{figure}
	\includegraphics[width=8.6cm]{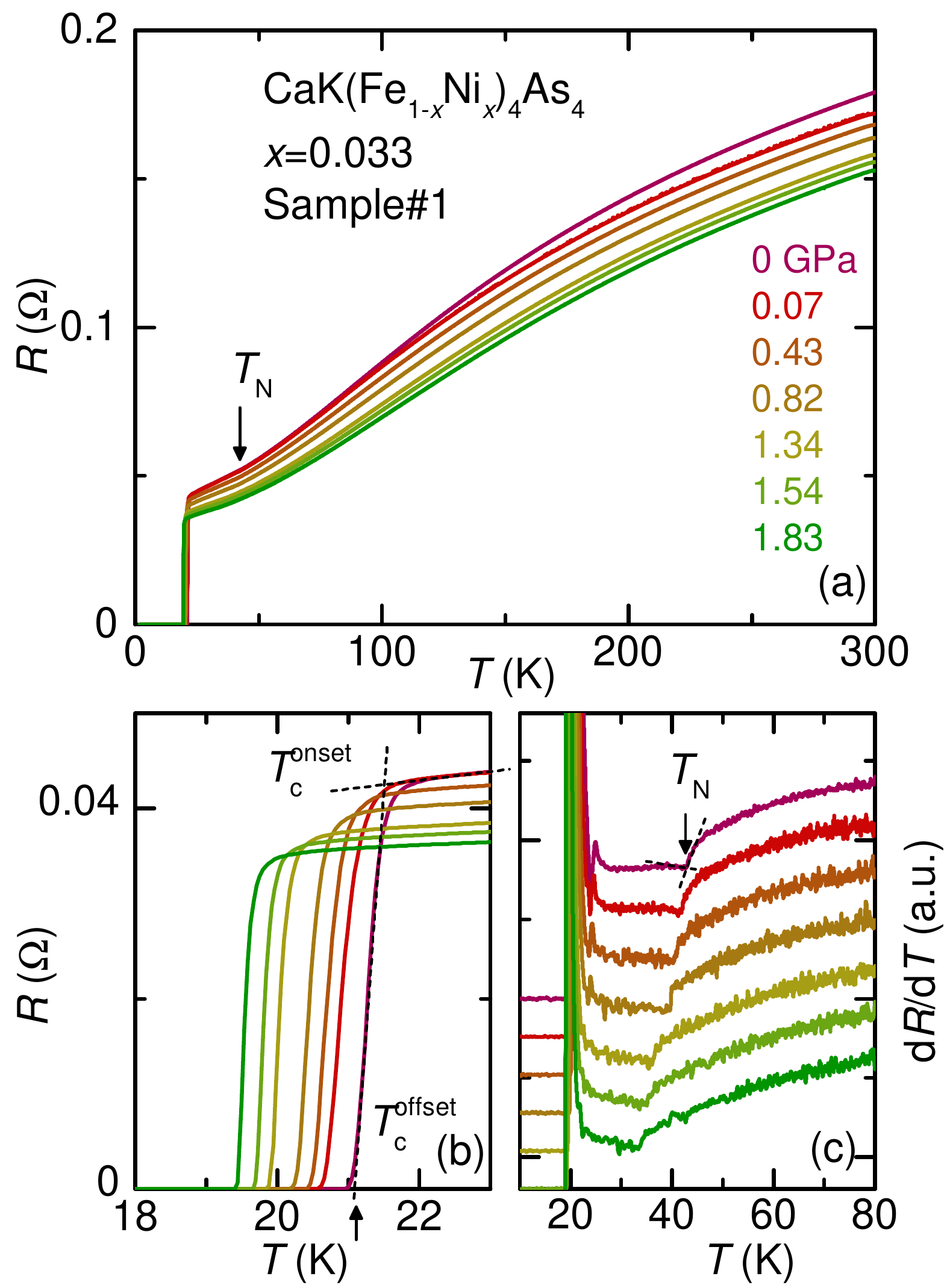}%
	\caption{(a) Evolution of the in-plane resistance with hydrostatic pressures up to 1.83 GPa measured in a PCC for the CaK(Fe$_{0.967}$Ni$_{0.033}$)$_4$As$_4$, sample\#1. (b) Blowup of the low temperature region. Criteria for $T_\textrm{c}^\textrm{onset}$ and $T_\textrm{c}^\textrm{offset}$ are indicated in the figure. (c) Temperature derivative, $dR/dT$, showing the evolution of the magnetic transition $T_\textrm{N}$ with offset criteria as shown in the figure.
		\label{lower_doping_piston}}
\end{figure}

\begin{figure}
	\includegraphics[width=8.6cm]{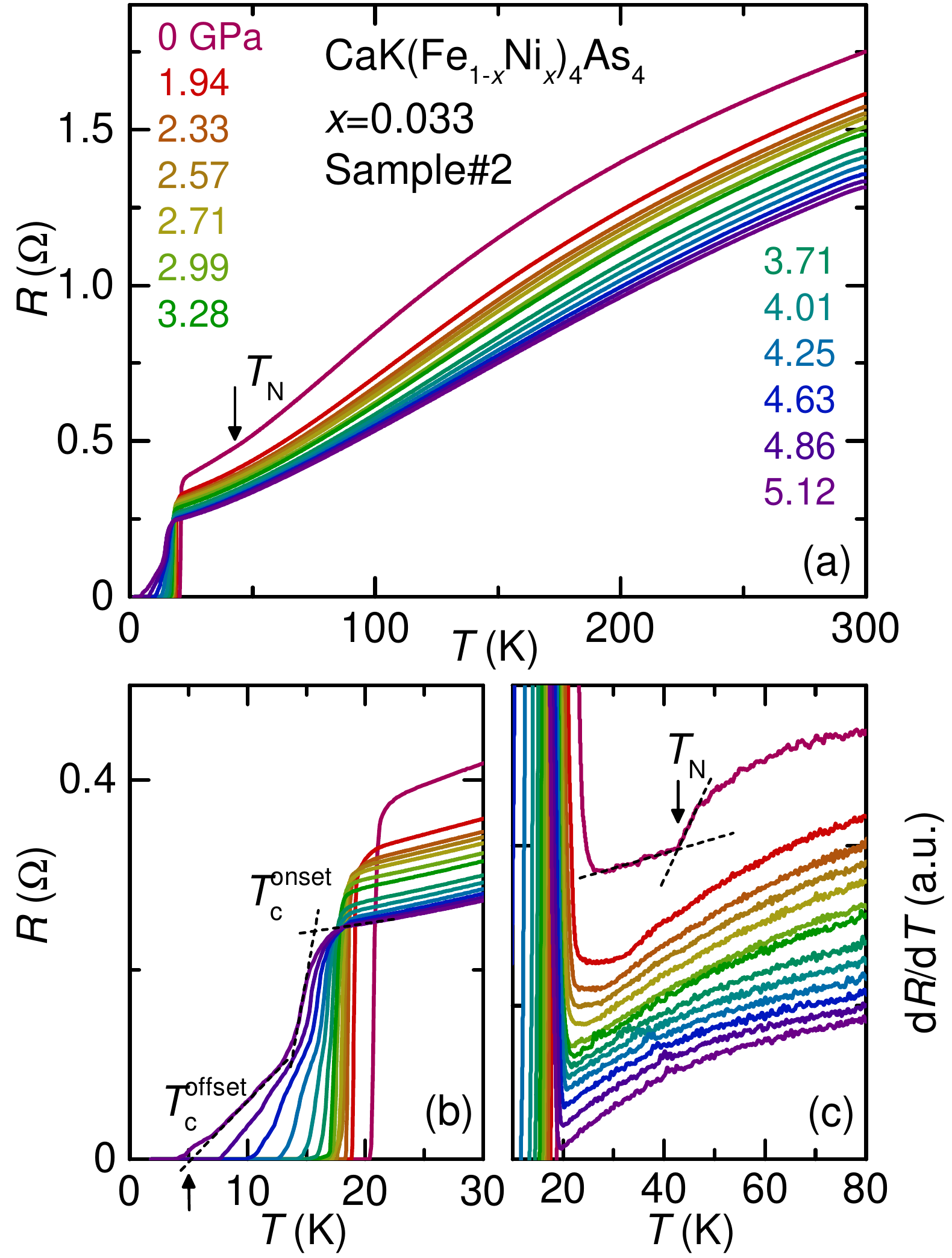}%
	\caption{(a) Evolution of the in-plane resistance with hydrostatic pressures up to 5.12 GPa measured in a mBAC for CaK(Fe$_{0.967}$Ni$_{0.033}$)$_4$As$_4$, sample\#2. (b) Blowup of the low temperature region. (c) Temperature derivative, $dR/dT$, showing the evolution of magnetic transition $T_\textrm{N}$.
		\label{lower_doping_mBAC}}
\end{figure}

\begin{figure}
	\includegraphics[width=8.6cm]{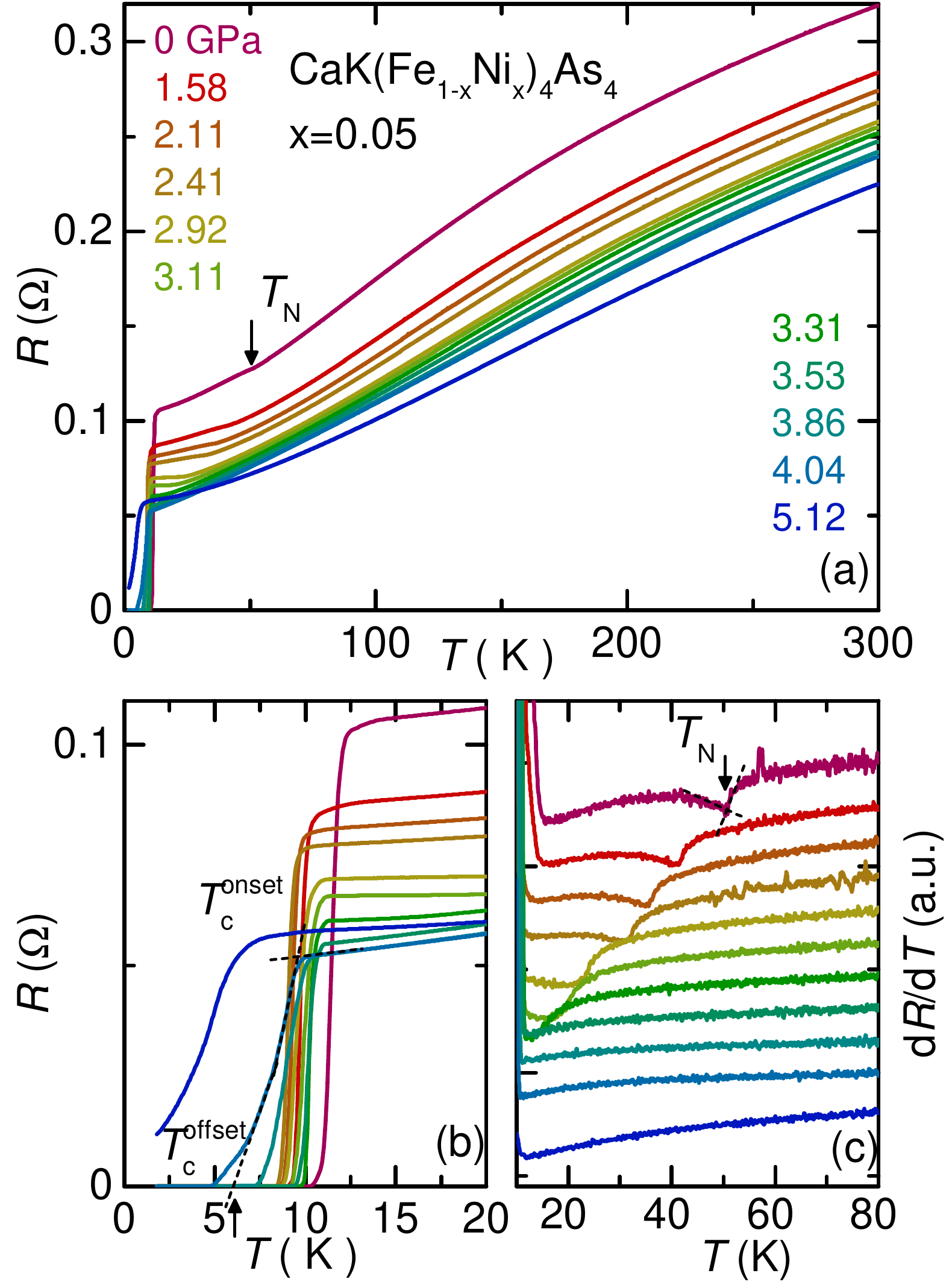}%
	\caption{(a) Evolution of the in-plane resistance with hydrostatic pressures up to 5.12 GPa measured in a mBAC for CaK(Fe$_{0.95}$Ni$_{0.05}$)$_4$As$_4$. (b) Blowup of the low temperature region. (c) Temperature derivative, $dR/dT$, showing the evolution of magnetic transition $T_\textrm{N}$.
		\label{higher_doping}}
\end{figure}

Figs. 1(a) and 2(a) show the pressure dependence of the temperature dependent resistance for CaK(Fe$_{1-x}$Ni$_{x}$)$_4$As$_4$, $x=0.033$. Sample $\#$1 was measured in the PCC for pressures up to 1.83 GPa. Sample $\#$2  was measured in the mBAC for pressures up to 5.12 GPa. For both samples, the 0 GPa resistance was corrected for geometric changes to the sample via normalization. (Details of the normalization are described in the Appendix.) Fig. 3(a) shows the pressure dependence of the temperature dependent resistance for the $x=0.050$ sample that was measured in the mBAC for pressures up to 5.12 GPa. In general, for all samples, the resistance decreases under applied pressure.

For both compositions, the magnetic phase transition $T_\textrm{N}$ appears as a kink-like anomaly in the lower temperature data and is more pronounced in the $x=0.050$ compound. This feature is more clearly revealed as a step-like anomaly in the temperature derivative $dR/dT$ [Figs. 1(c),2(c) and 3(c)]. These plots demonstrate that $T_\textrm{N}$ is suppressed by increasing pressure before it disappears at higher pressures.

\begin{figure}
	\includegraphics[width=8.6cm]{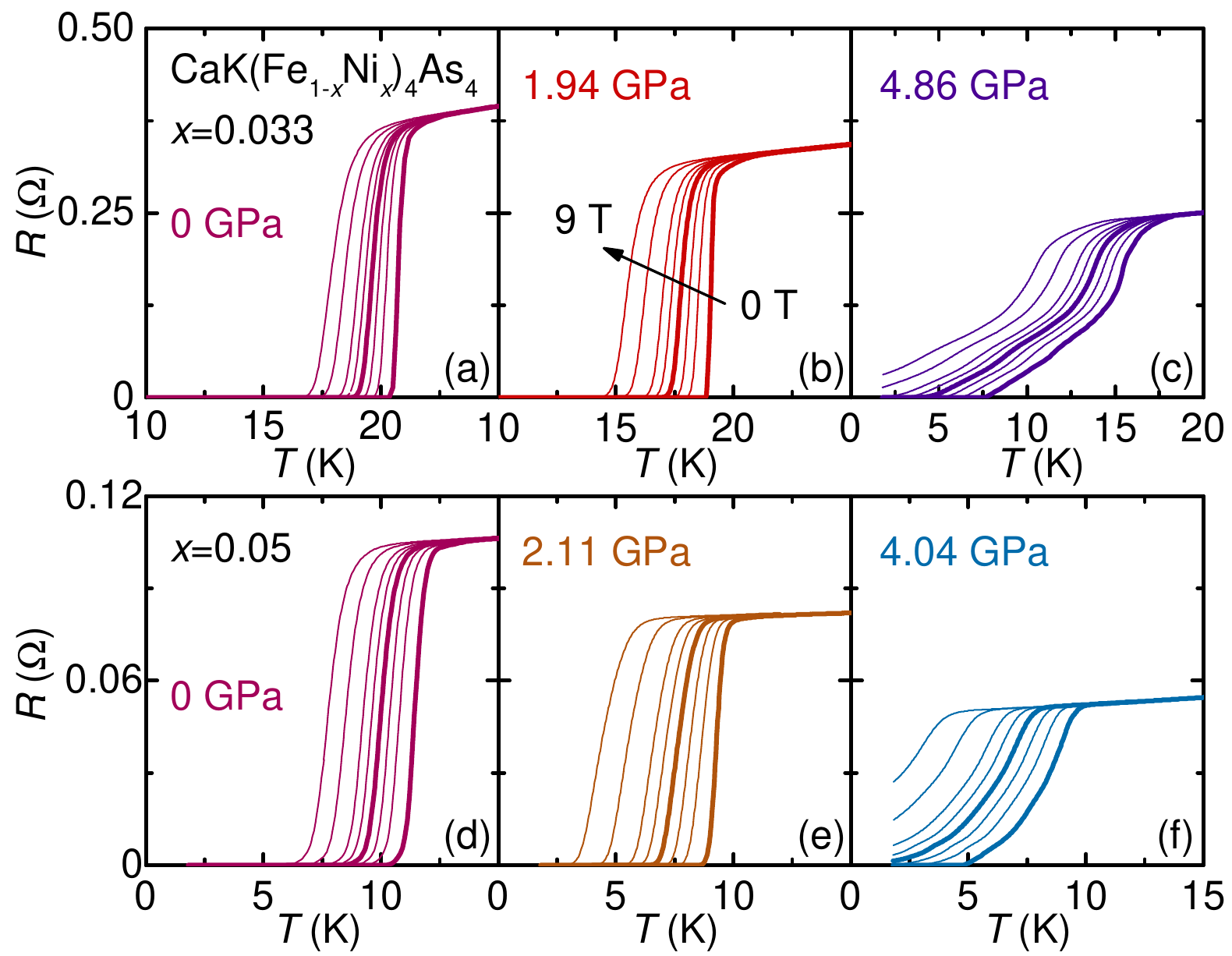}%
	\caption{Temperature dependence of resistance under magnetic field up to 9 T for selective pressures for CaK(Fe$_{1-x}$Ni$_{x}$)$_4$As$_4$, $x=0.033$ ((a)-(c)), $x=0.050$ ((d)-(f)). Superconducting transition becomes broader as pressure is increased for both compounds, to explore the nature of the broadening, transition width at 0 T and 3 T (indicated by thick lines in the figures) were analyzed and described in details in the text. 
		\label{RT_H}}
\end{figure}

\begin{figure}
	\includegraphics[width=8.6cm]{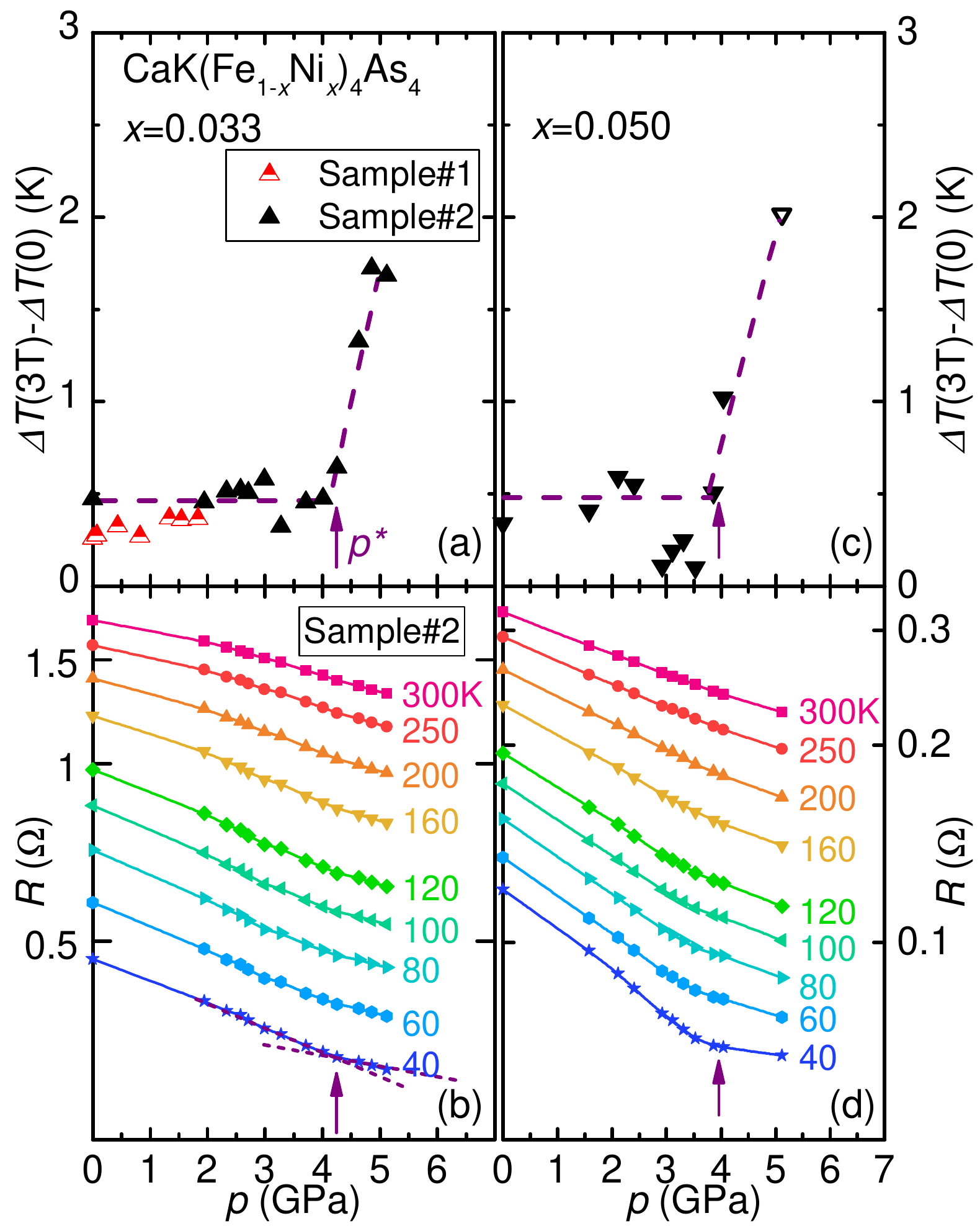}%
	\caption{(a),(c) Pressure dependence of the superconducting transition widths difference  for CaK(Fe$_{1-x}$Ni$_{x}$)$_4$As$_4$, $x=0.033$ and $0.050$ respectively. The superconducting transition widths is $\Delta T=T_\textrm{c}^\textrm{onset}-T_\textrm{c}^\textrm{offset}$ and the widths difference is taken between 0 field and 3 T. Open symbol in panel (c) is the widths difference taken between 0 field and 1 T because of no clear definition of $T_\textrm{c}^\textrm{offset}$ at 3 T for 5.12 GPa. (b), (d) Pressure dependence of resistance at $R(p)$ fixed temperatures for CaK(Fe$_{1-x}$Ni$_{x}$)$_4$As$_4$, $x=0.033$ and $0.050$ respectively. The critical pressure $p^*$(Arrows in the figure) which is associated with the hcT phase is described in details in the text.
		\label{HT_phase}}
\end{figure}

The blowups of the low temperature resistance [Figs. 1(b), 2(b) and 3(b)] show how $T_\textrm{c}$ changes under increasing pressure. For $x=0.033$, $T_\textrm{c}$ monotonically decreases in the studied pressure range. In contrast, for $x=0.050$, after 2.41 GPa there is a slight enhancement of the $T_\textrm{c}$ before it is suppressed again at higher pressures. 

Upon increasing pressures above $\sim$ 4 GPa, the sharp superconducting transition at lower pressures becomes broadened at higher pressures. A similar behavior was also observed in the parent compound CaKFe$_4$As$_4$ and has been associated with the hcT phase transition at $p \gtrsim$ 4 GPa\cite{Kaluarachchi2017PRB}. In order to understand the nature of the broadening in the substituted system, analysis similar to that in Ref. \onlinecite{Kaluarachchi2017PRB} was carried out.

Fig. \ref{RT_H} presents the temperature dependence of the resistance under magnetic field up to 9 T for selected pressures. The superconducting transition width, $\Delta T=T_\textrm{c}^\textrm{onset}-T_\textrm{c}^\textrm{offset}$, is broadened with increasing pressure, with the criteria for $T_\textrm{c}^\textrm{onset}$ and $T_\textrm{c}^\textrm{offset}$ shown in Figs. 1(b), 2(b) and 3(b). In order to determine whether the broadening is associated with any sort of phase transition, or is simply due to pressure inhomogeneities in the pressure medium when larger loads are applied, the field dependence of the superconducting transition width $\Delta T(H)$ was studied\cite{Kaluarachchi2017PRB}. Specifically, the transition width at magnetic fields 0 T and at 3 T (indicated by thicker lines in Figs. \ref{RT_H}) were determined, and then the difference between them, $\Delta T(3\textrm T)-\Delta T(0)$, was calculated. Any broadening due to the pressure inhomogeneities are expected to be equally present in the $H=0$ T and 3 T data. Figs. \ref{HT_phase}(a) and (c) present the pressure dependence of the transition width difference. As it is clearly shown, for both compositions, $\Delta T(3\textrm T)-\Delta T(0)$ increases dramatically as pressure goes above $p^* \sim$ 4 GPa (indicated by arrows in Figs. \ref{HT_phase}(a), (c)). Note that for $x=0.050$, at 5.12 GPa, the transition width difference was taken between $H=0$ T and 1 T, because $T_\textrm{c}^\textrm{offset}$ is not clearly defined at $H=3$T. But we would expect the transition width difference between $H=0$ T and 3 T to be even larger at this pressure. Furthermore, the pressure dependence of the resistance $R(p)$ at fixed temperatures for both compositions (Figs. \ref{HT_phase}(b), (d)) shows anomaly at the same pressure at 40 K (indicated by arrows in the figure), though subtle for $x=0.033$. Based on the analogy with the parent compound CaKFe$_4$As$_4$\cite{Kaluarachchi2017PRB}, we identify this anomaly as an indication of the hcT phase transition that exists from base temperature up to at least 40 K. As was the case for pure CaKFe$_4$As$_4$, we believe that superconductivity is not bulk for $p \gtrsim$ 4 GPa (i.e. in the hcT phase).


\begin{figure}
	\includegraphics[width=8.6cm]{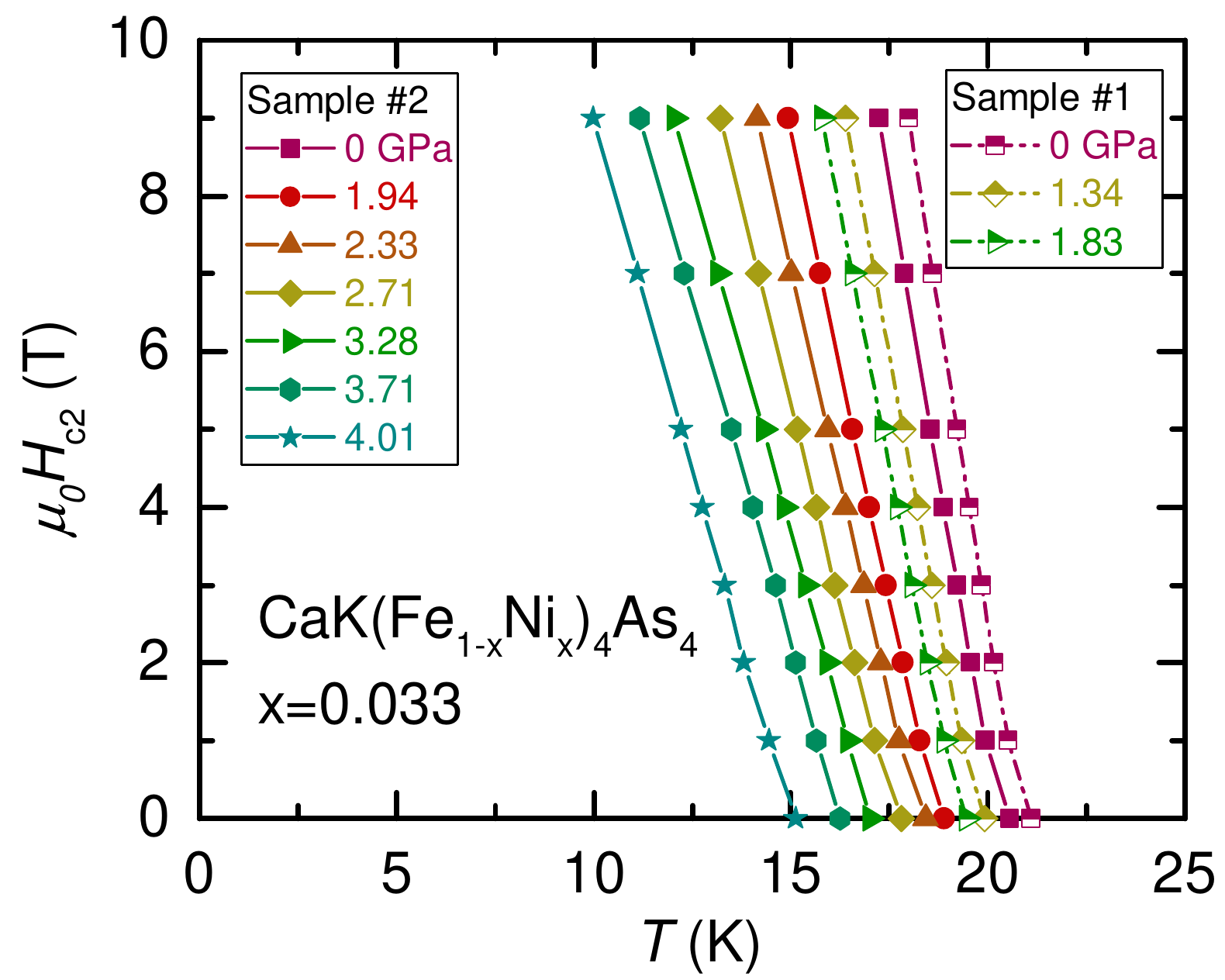}%
	\caption{Temperature dependence of the upper superconducting critical field, $H_{\textrm c2}(T)$, under selected pressures for CaK(Fe$_{1-x}$Ni$_{x}$)$_4$As$_4$, $x=0.033$. $T_\textrm{c}^\textrm{offset}$ is used for the figure. Half filled and solid symbols are two samples measured in PCC and mBAC respectively.
		\label{Hc2_lowerdoping}}
\end{figure}

\begin{figure}
	\includegraphics[width=8.6cm]{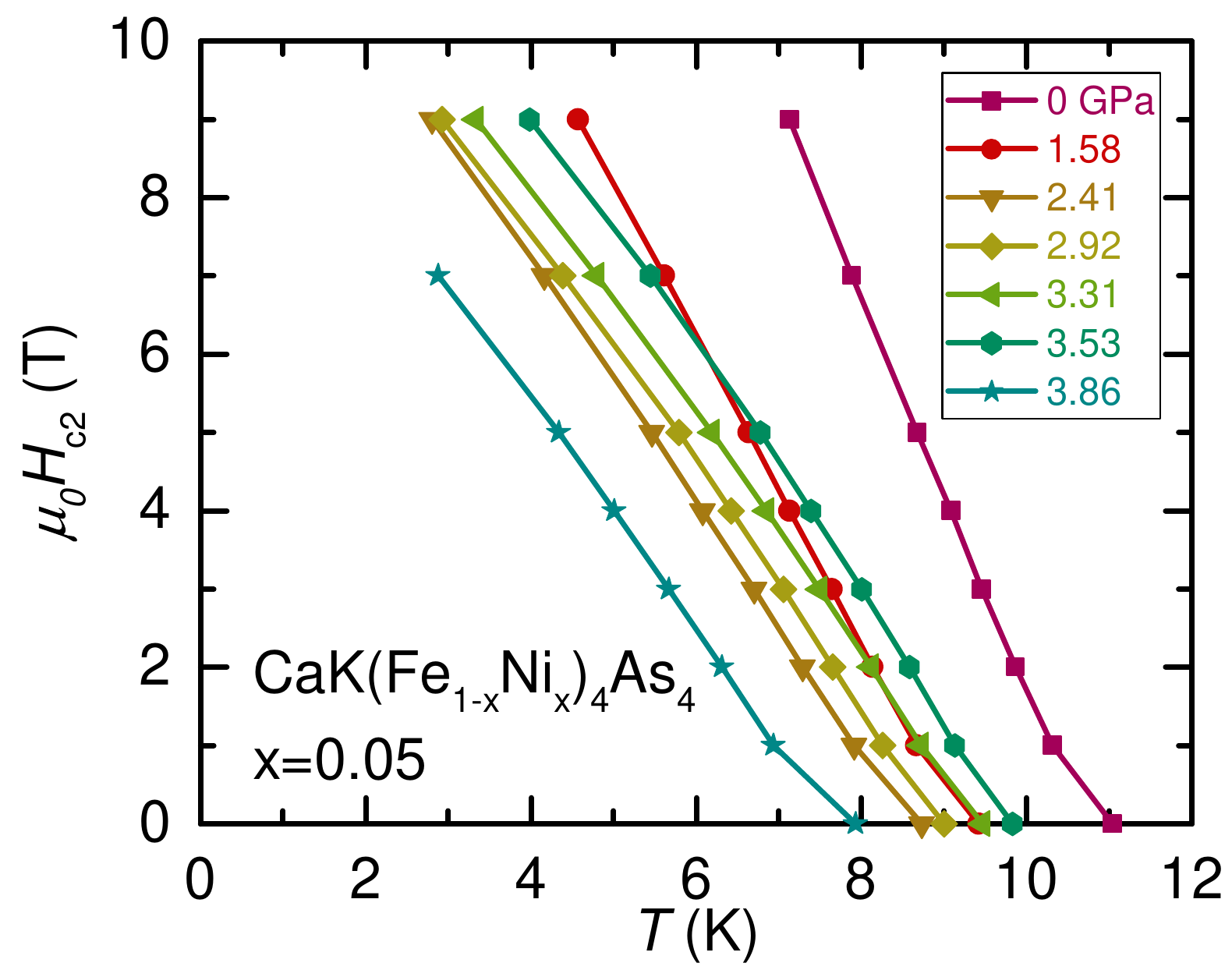}%
	\caption{Temperature dependence of the upper superconducting critical field, $H_{\textrm c2}(T)$, under selected pressures for CaK(Fe$_{1-x}$Ni$_{x}$)$_4$As$_4$, $x=0.050$. $T_\textrm{c}^\textrm{offset}$ is used for the figure.
		\label{Hc2_higherdoping}}
\end{figure}

\begin{figure}
	\includegraphics[width=8.6cm]{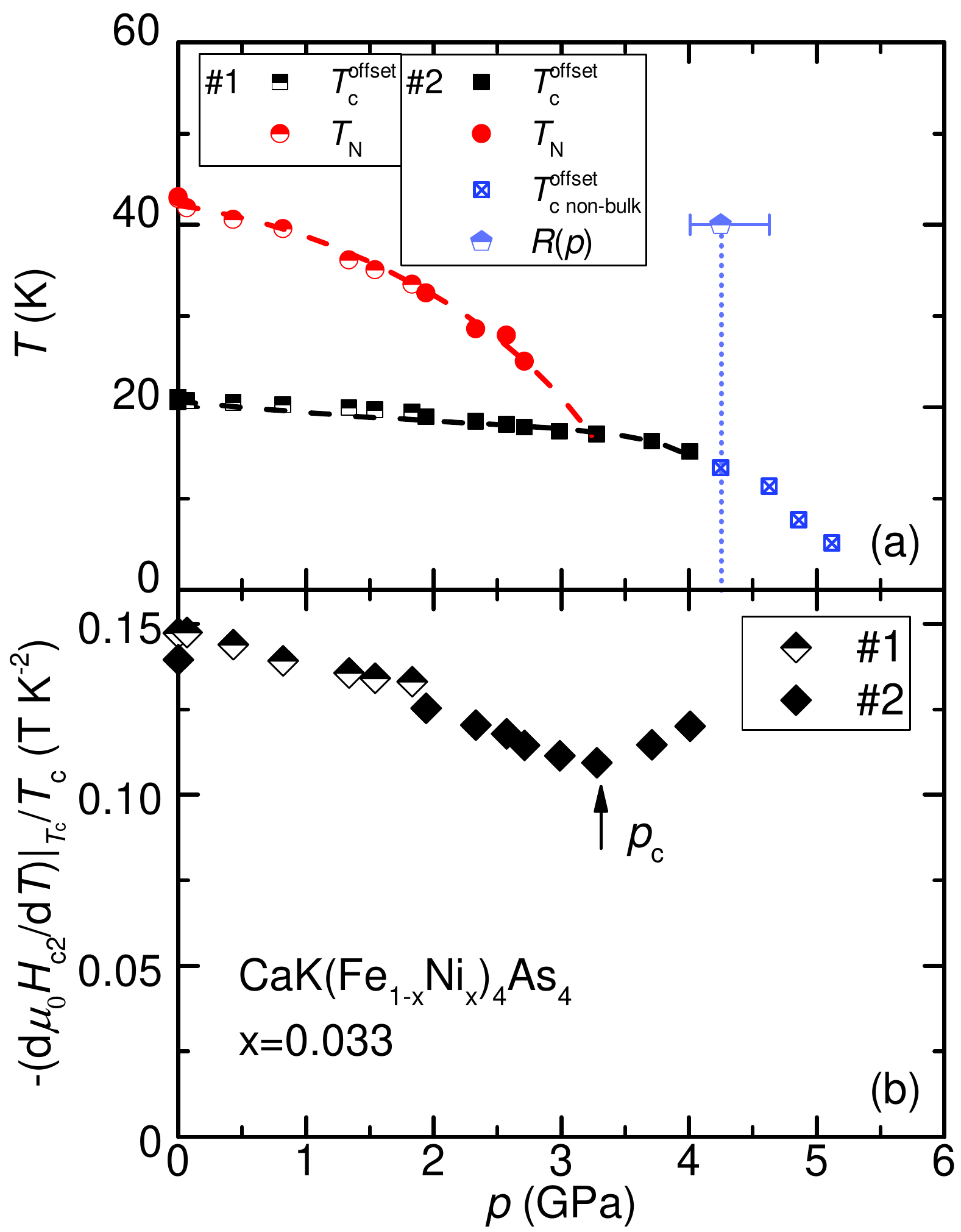}%
	\caption{(a) Temperature-pressure phase diagram of CaK(Fe$_{1-x}$Ni$_{x}$)$_4$As$_4$, $x=0.033$, as determined from resistance measurement. The squares and circles represent the superconducting $T_\textrm{c}^\textrm{offset}$ and magnetic $T_\textrm{N}$ phase transition. Half filled and solid symbols are two samples measured in the PCC and the mBAC respectively. Blue symbols represent $T_\textrm{c}^\textrm{offset}$ for filamentary superconductivity. Dashed lines are guides to the eye. Blue dotted line indicates the half-collapsed-tetragonal phase transition up to 40 K, inferred from the pressure dependent resistance $R(p)$ data in Fig. \ref{HT_phase}. (b) Pressure dependence of the normalized upper critical field slope -(1/$T_\textrm c$)($d\mu_oH_{\textrm c2}$/$dT$)$|_{T_\textrm c}$. A local minimum in the slope at $p_c$ (indicated by arrow) is observed near the pressure where $T_\textrm{c}^\textrm{offset}$ and $T_\textrm{N}$ lines cross.
		\label{phase_diagram_lowerdoping}}
\end{figure}

\begin{figure}
	\includegraphics[width=8.6cm]{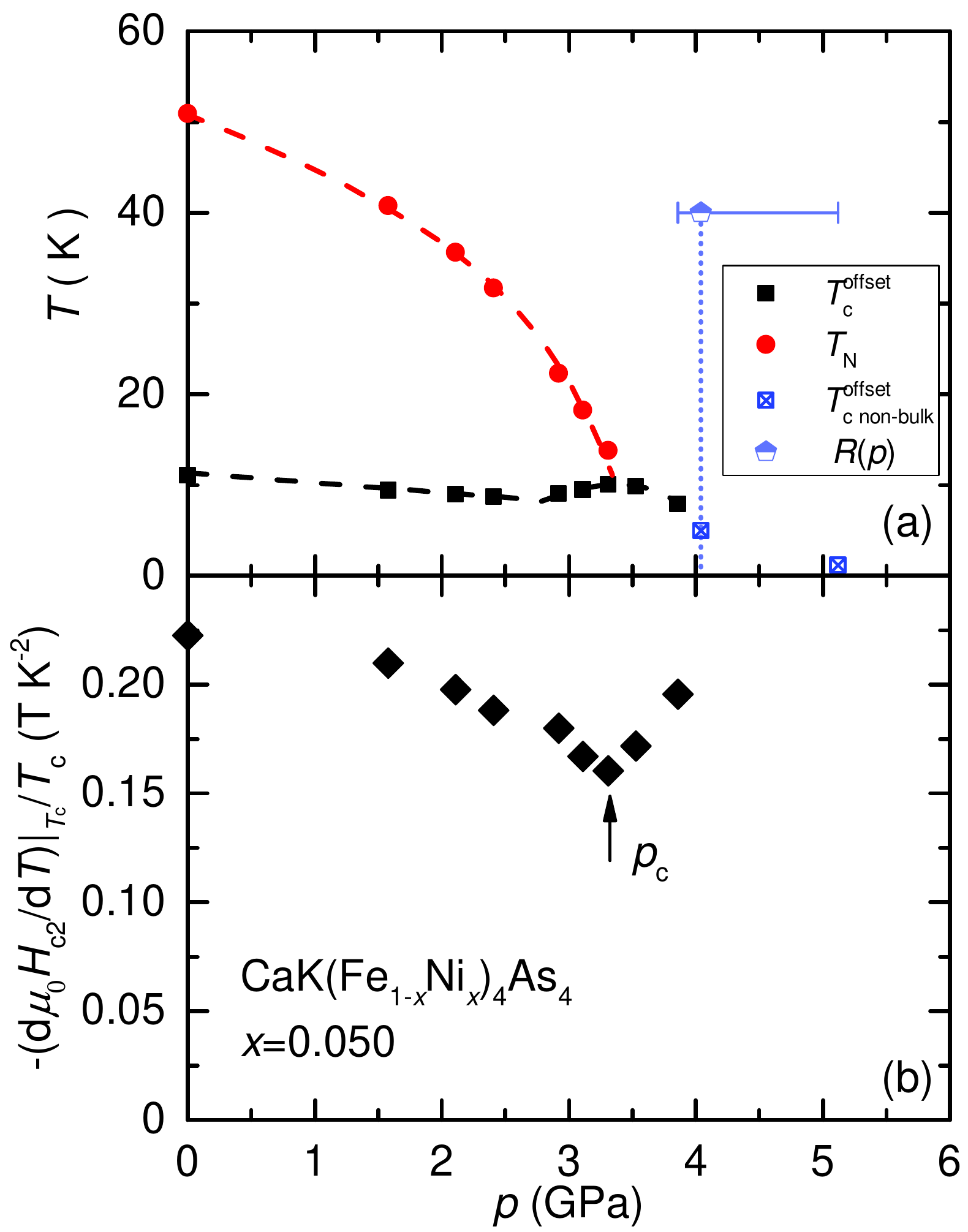}%
	\caption{(a) Temperature-pressure phase diagram of CaK(Fe$_{1-x}$Ni$_{x}$)$_4$As$_4$, $x=0.050$, as determined from resistance measurement. The squares and circles represent the superconducting $T_\textrm{c}^\textrm{offset}$ and magnetic $T_\textrm{N}$ phase transition. Blue symbols represent $T_\textrm{c}^\textrm{offset}$ for filamentary superconductivity. Dashed lines are guides to the eye. Blue dotted line indicates the half-collapsed-tetragonal phase transition up to 40 K, inferred from the pressure dependent resistance $R(p)$ data in Fig. \ref{HT_phase}. (b) Pressure dependence of the normalized upper critical field slope -(1/$T_\textrm c$)($d\mu_oH_{\textrm c2}$/$dT$)$|_{T_\textrm c}$. A local minimum in the slope at $p_c$ (indicated by arrow) is observed near the pressure where $T_\textrm{c}^\textrm{offset}$ and $T_\textrm{N}$ lines cross.
		\label{phase_diagram_higherdoping}}
\end{figure}

The upper superconducting critical field $H_\textrm{c2}$ can be evaluated from Fig. \ref{RT_H} at pressures lower than $p^*$, where superconductivity is considered bulk, using the offset criteria defined in Figs. 1-3. The temperature dependence of $H_\textrm{c2}$ at various pressures is presented in Figs. \ref{Hc2_lowerdoping} and \ref{Hc2_higherdoping} for  CaK(Fe$_{1-x}$Ni$_{x}$)$_4$As$_4$, $x=0.033$ and 0.050 respectively. For $x=0.033$, both Sample$\#$1 and Sample$\#$2 were analyzed and plotted in Fig \ref{Hc2_lowerdoping}. Note that at ambient pressure, $T_\textrm{c}^\textrm{offset}$ values for two samples differ by $\sim$0.5 K, possibly due to a small difference of the substitution level at different positions of the crystal they were cut from. As is shown in  Figs. \ref{Hc2_lowerdoping} and \ref{Hc2_higherdoping}, for $x=0.033$, $H_\textrm{c2}$ is systematically suppressed by increasing pressure, whereas, for $x=0.050$, the evolution of the temperature dependent $H_\textrm{c2}$ is nonmonotonic. For both compositions, $H_\textrm{c2}$ is linear in temperature except for magnetic fields below 1 T. The curvature at low fields has been observed in other FeSC and can be explained by nature of superconductivity\cite{Kogan2012,Kaluarachchi2016,Xiang2017PRB}, which is also the case for the parent compound CaKFe$_4$As$_4$\cite{Mou2016PRL}.

Figs. $\ref{phase_diagram_lowerdoping}$(a) and $\ref{phase_diagram_higherdoping}$(a) present the $p-T$ phase diagrams for CaK(Fe$_{1-x}$Ni$_{x}$)$_4$As$_4$, $x=0.033$ and 0.050 respectively, with $T_\textrm{c}^\textrm{offset}$ and $T_\textrm{N}$ values obtained using the criteria shown in Figs. 1-3 and the indication of non-bulk superconductivity above $p^*$. For both compositions, $T_\textrm{N}$ is suppressed by pressure, specifically, $T_\textrm{N}$ is suppressed from 43 K to 25 K at 2.71 GPa for $x=0.033$ and suppressed from 51 K to 13.8 K at 3.31 GPa for $x=0.050$.

In terms of superconductivity, for x=0.033, $T_\textrm{c}^\textrm{offset}$ is monotonically suppressed with increasing pressure. It drops from 20.5 K to 15.1 K at 4.01 GPa before superconductivity becomes non-bulk. A closer examination reveals that $T_\textrm{c}^\textrm{offset}$ is initially linearly suppressed by pressure up to 2.71 GPa, then a small, but clear deviation from the linear suppression was observed above 2.99 GPa. An extrapolation of $T_\textrm{N}$ shows that the deviation happens near the crossing of $T_\textrm{N}$ and $T_\textrm{c}^\textrm{offset}$ lines. For $x=0.050$, the behavior of $T_\textrm{c}^\textrm{offset}(p)$ is distinctly non-monotonic. $T_\textrm{c}^\textrm{offset}$ is initially linearly suppressed from 11 K to a local minimum of 8.7 K at 2.41 GPa. Then it rises to a maximum of 10 K at 3.31 GPa, exhibiting a dome shape. This dome of enhanced $T_\textrm{c}^\textrm{offset}$ coincides with the disappearance of $T_\textrm{N}$. After the local maximum in $T_\textrm{c}^\textrm{offset}$ there is a much more rapid suppression of $T_\textrm{c}^\textrm{offset}$ with increasing $p$ until the hcT transition at $p^*$. For both compositions, a change in $T_\textrm{c}^\textrm{offset}(p)$ happens at the pressure where $T_\textrm{N}$ and $T_\textrm{c}^\textrm{offset}$ lines cross.


Both compositions show signatures of non-bulk superconductivity above $p^* \sim$ 4 GPa (blue symbols in Figs. \ref{phase_diagram_lowerdoping}(a), \ref{phase_diagram_higherdoping}(a)) similar to the parent compound CaKFe$_4$As$_4$\cite{Kaluarachchi2017PRB}, suggesting the same hcT phase transition. Pressure dependent resistance data in Fig. \ref{HT_phase} demonstrates that the hcT phase transition is discernable up to at least 40 K for the substituted compounds. The transition pressure does not appear to change with Ni-substitution. This is not too surprising given the fact that the hcT transition does not involve the Fe-plane but is, instead As-As bonding across the Ca plane.


To better understand the superconducting properties of CaK(Fe$_{1-x}$Ni$_{x}$)$_4$As$_4$, the superconducting upper critical field $H_\textrm{c2}$ was analyzed following Refs.\onlinecite{Taufour2014,Kaluarachchi2016,Xiang2017PRB}. Generally speaking, the slope of the upper critical field normalized by $T_\textrm{c}$, is related to the Fermi velocity and superconducting gap of the system\cite{Kogan2012}. In the clean limit, for a single-band,
\begin{equation}
-(1/T_\textrm c)(d\mu_oH_{\textrm c2}/dT)|_{T_\textrm c} \propto 1/v_F^2,
\label{eq:Hc2}
\end{equation}
where $v_F$ is the Fermi velocity. Even though the superconductivity in CaKFe$_4$As$_4$ compounds is multiband, Eq. \ref{eq:Hc2} can give qualitative insight into changes induced by pressure.

As is shown in Figs. $\ref{phase_diagram_lowerdoping}$(b) and $\ref{phase_diagram_higherdoping}$(b), the normalized slope of the upper critical field -(1/$T_\textrm c$)($d\mu_oH_{\textrm c2}$/$dT$)$|_{T_\textrm c}$ (the slope $d\mu_oH_{\textrm c2}$/$dT$$|_{T_\textrm c}$) is calculated by linear fitting the data from 1-5 T in Figs. $\ref{Hc2_lowerdoping}$ and $\ref{Hc2_higherdoping}$) exhibits a similar pressure dependence for $x=0.033$ and 0.050. It initially decreases upon increasing pressure and then begins to increase above pressure $p_c$, resulting in a minimum of -(1/$T_\textrm c$)($d\mu_oH_{\textrm c2}$/$dT$)$|_{T_\textrm c}$ in the studied pressure range. In both compositions, $p_c$ coincides with the crossing of $T_\textrm{N}$ and $T_\textrm{c}^\textrm{offset}$ lines, suggesting a common origin of this feature. 

In Fe-based superconductors, especially the "122" system, Fermi-surface nesting can lead to a partial opening of a gap at the Fermi-surface below $T_\textrm{N}$. By tuning with doping or applying pressure, a Fermi-surface reconstruction could happen due to the disappearance of magnetism\cite{Jiang2009,Dai2009,Gooch2009PRB,Gooch2010,Maiwald2012PRB,Arsenijevic2013PRB,Liu2009PRL,Liu2009PRB,Dhaka2013PRL}. For CaK(Fe$_{1-x}$Ni$_{x}$)$_4$As$_4$ ($x=0.033$ and $0.050$), a clear change of the pressure dependence of the normalized slope -(1/$T_\textrm c$)($d\mu_oH_{\textrm c2}$/$dT$)$|_{T_\textrm c}$ is observed at $p_c$, indicating a possible Fermi-surface reconstruction near $p_c$. Note that for $x=0.050$, there appears to be a discontinuous change in the normalized slope -(1/$T_\textrm c$)($d\mu_oH_{\textrm c2}$/$dT$)$|_{T_\textrm c}$ and a subtle anomaly in $T_\textrm{c}(p)$ from 2.41 GPa to 2.92 GP, suggesting there may be a Liftshiz transition near this pressure. Such features are not observed for $x=0.033$.

Figs. \ref{phase_diagram_lowerdoping} and \ref{phase_diagram_higherdoping}, then, combine surprising and not unexpected features. The hcT phase transition pressure appears insensitive to Ni subsititution. This is reasonable because this transition involves bonding of As atoms across the Ca-plane. The clear feature at $p_c$ in -(1/$T_\textrm c$)($d\mu_oH_{\textrm c2}$/$dT$)$|_{T_\textrm c}$, as well as the more subtle features in $T_\textrm{c}(p)$, are again not too surprising and can be associated with the change (with increasing $p$) from $T_\textrm{N} > T_\textrm{c}$ to $T_\textrm{N} < T_\textrm{c}$, i.e. $T_\textrm{c}$ occurring in an AFM ordered state to $T_\textrm{c}$ occurring in a state lacking the AFM order and associated additional periodicities. The surprising feature shown in Figs. \ref{phase_diagram_lowerdoping} and \ref{phase_diagram_higherdoping} is the weak suppression of $T_\textrm{c}$ concurrent with the strong suppression of $T_\textrm{N}$. This is contrary to what is seen in Co substitution and pressure study on BaFe$_2$As$_2$(where $T_\textrm{c}$ increases, as $T_\textrm{N}$ is suppressed)\cite{Colombier2009,Colombier2010,Canfield2010,Ni2011} and brings into question the exact effects suppression of $T_\textrm{N}$ has on the magnetic fluctuations that the superconducting state is nominally built out of.

\section{Conclusion}

In conclusion, the resistance of Ni-substituted iron-based superconductor CaK(Fe$_{1-x}$Ni$_{x}$)$_4$As$_4$ ($x=0.033$ and $0.050$) has been studied under pressures up to 5.12 GPa and in magnetic fields up to 9 T. For both substitution levels, hedgehog spin-vortex-crystal magnetic transition temperature, $T_\textrm{N}$, is suppressed with increasing pressure. In both compositions, $T_\textrm{c}$ is initially suppressed as well and exhibits a weak anomaly near the crossing of $T_\textrm{N}$ and $T_\textrm{c}$ lines. As pressure exceeds $\sim$ 4 GPa, both compositions likely go through the half-collapsed-tetragonal phase transition, similar to the one observed in the parent compound. This demonstrates the insensitivity of the hcT transition pressure to Ni-substitution. The minimum observed in the normalized slope of the upper critical field, -(1/$T_\textrm c$)($d\mu_oH_{\textrm c2}$/$dT$)$|_{T_\textrm c}$, at the pressure where $T_\textrm{N}$ and $T_\textrm{c}$ lines cross indicate a possible Fermi-surface reconstruction associated with the disappearance of antiferromagnetism.

\begin{acknowledgements}
We would like to thank A. Kreyssig for useful discussions. This work is supported by the US DOE, Basic Energy Sciences, Materials Science and Engineering Division under contract No. DE-AC02-07CH11358. L. X. was supported, in part, by the W. M. Keck Foundation. W. R. M. was supported by the Gordon and Betty Moore Foundation's EPiQS Initiative through Grant GBMF4411.
\end{acknowledgements}

\section{APPENDIX}

\begin{figure}
	\includegraphics[width=8.6cm]{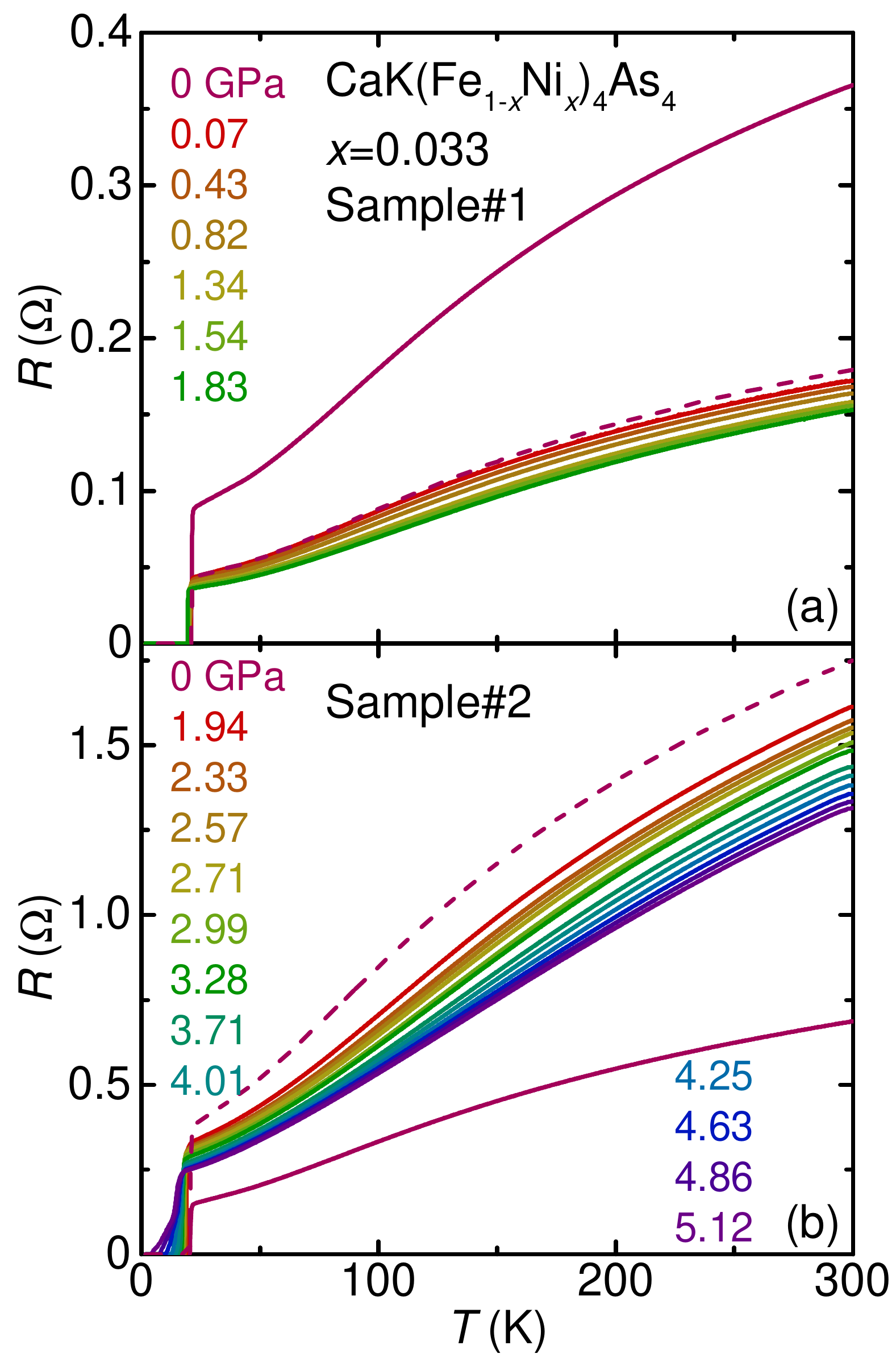}%
	\caption{Evolution of the in-plane resistance with hydrostatic pressure of Sample\#1 measured in a PCC (a) and Sample\#2 measured in a mBAC (b) for CaK(Fe$_{1-x}$Ni$_{x}$)$_4$As$_4$, $x=0.033$. Solid lines are the actual resistance data measured, dashed lines are the normalized resistance for 0 GPa. Notice that the 0 GPa resistance is measured on PPMS puck outside of either pressure cell (i.e. ambient pressure); in both cases there is a sudden change between the resistance measured at ambient pressure and inside pressure cell. Possible reasons for the sudden change and details of normalization are explained in details in the text.
		\label{RT}}
\end{figure}

\begin{figure}
	\includegraphics[width=8.6cm]{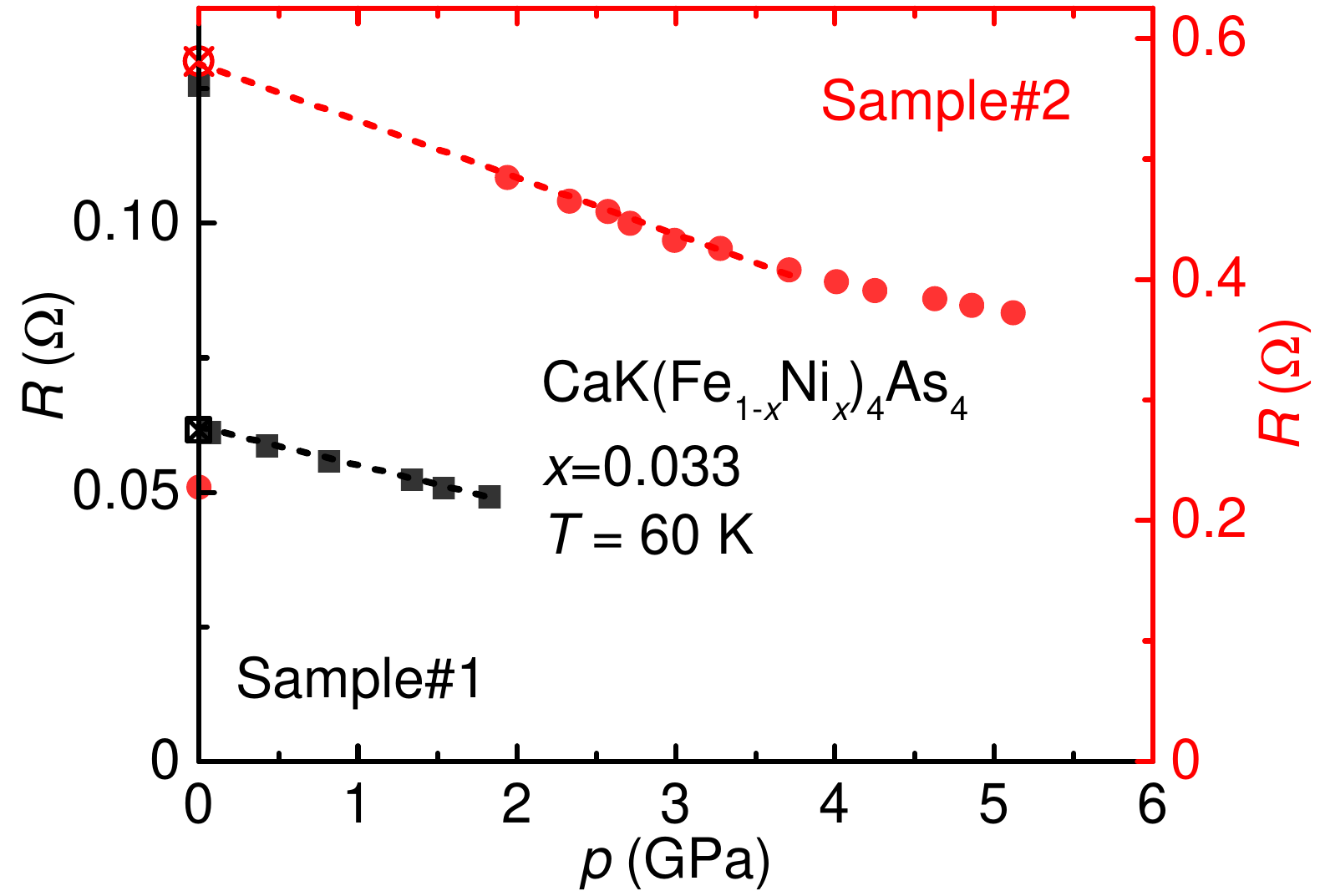}%
	\caption{Pressure dependence of resistance at 60 K for CaK(Fe$_{1-x}$Ni$_{x}$)$_4$As$_4$, $x=0.033$, black solid squares are data from Sample\#1 measured in PCC, red solid circles are data from Sample\#2 measured in mBAC. Dashed lines are linear fitting of the data before 4 GPa (not including 0 GPa), notice the clear deviation from the linear fitting for the 0 GPa data. Open symbols are the corresponding normalized 0 GPa resistance for Sample\#1 and Sample\#2 at 60 K.
		\label{RP_60K}}
\end{figure}

Fig. \ref{RT} presents the evolution of the in-plane resistance with hydrostatic pressure for CaK(Fe$_{1-x}$Ni$_{x}$)$_4$As$_4$, $x=0.033$, solid lines in the figure are the actual measured resistance data, dashed lines are the resistance after normalization. Sample\#1 was measured in a PCC for pressures up to 1.83 GPa and Sample\#2 was measured in a mBAC for pressures up to 5.12 GPa. Note that the 0 GPa resistance data was measured on a PPMS puck outside of either pressure cell (i.e. ambient pressure), a sudden change of resistance between ambient pressure and inside pressure cell was observed in both samples. For Sample\#1, when the sample was moved from PPMS puck and mounted onto the PCC, one contact of the voltage channel became detached from the sample and that contact had to be re-attached. As a result, the changed position of the contact led to changes in the resistance before and after. For Sample\#2, nothing was intentionally done to the sample before and after it was put into the mBAC, the sudden change of the resistance is most likely due to the exfoliation or cracking of the sample when pressure was first applied as the pressure cell was closed. Despite the abrupt change of resistance from ambient pressure to the first finite pressures inside the pressure cell, the resistance of CaK(Fe$_{1-x}$Ni$_{x}$)$_4$As$_4$ ($x=0.033$ and $0.05$) continuously and systematically decreases upon increasing pressure, consistent with the behavior that is observed in parent compound CaKFe$_4$As$_4$\cite{Meier2018} and many "122" systems\cite{Hassinger2012PRB,Hassinger2016PRB,Taufour2014}.

To better evaluate the resistance evolution with pressure, especially the pressure dependence of resistance at various temperatures (Fig. \ref{HT_phase} (b)(d)), the ambient pressure resistance is shifted via normalization (assuming in each case that the shift was due to geometric changes). Fig. \ref{RP_60K} presents the pressure dependence of the resistance at $T=60$ K for Sample\#1 and Sample\#2 (solid symbols). Note $T=60$K was chosen because the pressure values are determined from the the $T_\textrm{c}(p)$ of lead\cite{Bireckoven1988} at $\sim$7 K, and the pressure cells are known to have pressure changes with temperature. With the pressure cells and liquid medium we used in this study, the pressure change from room temperature to 7 K can be $0.2\sim0.3$ GPa\cite{Thompson1984,Colombier2007}. 60 K was chosen based on the idea that at this temperature, the pressure medium has already solidified\cite{Torikachvili2015}, the temperature dependence of the thermal expansion of cell materials flattens at low temperature, and the pressure difference between 60 K and 7 K should be small\cite{Thompson1984}. The fact that 60 K is still above the magnetic transition temperature $T_\textrm{N}$ guarantees that pressure dependence of resistance at this temperature gives no feature related to magnetism. As shown in Fig. \ref{RP_60K}, except the ambient pressure data, the 60 K resistance for both samples are linearly suppressed by pressure before 4 GPa, so it is assumed that the ambient pressure resistance should also follow this pressure dependence (open symbols in Fig. \ref{RP_60K}). To do that, the ambient pressure resistance curves for the two samples are multiplied by two corresponding factors and moved to the dashed lines as shown in Fig. \ref{RT}.

\clearpage

\bibliographystyle{apsrev4-1}

\begin{thebibliography}{50}%
	\makeatletter
	\providecommand \@ifxundefined [1]{%
		\@ifx{#1\undefined}
	}%
	\providecommand \@ifnum [1]{%
		\ifnum #1\expandafter \@firstoftwo
		\else \expandafter \@secondoftwo
		\fi
	}%
	\providecommand \@ifx [1]{%
		\ifx #1\expandafter \@firstoftwo
		\else \expandafter \@secondoftwo
		\fi
	}%
	\providecommand \natexlab [1]{#1}%
	\providecommand \enquote  [1]{``#1''}%
	\providecommand \bibnamefont  [1]{#1}%
	\providecommand \bibfnamefont [1]{#1}%
	\providecommand \citenamefont [1]{#1}%
	\providecommand \href@noop [0]{\@secondoftwo}%
	\providecommand \href [0]{\begingroup \@sanitize@url \@href}%
	\providecommand \@href[1]{\@@startlink{#1}\@@href}%
	\providecommand \@@href[1]{\endgroup#1\@@endlink}%
	\providecommand \@sanitize@url [0]{\catcode `\\12\catcode `\$12\catcode
		`\&12\catcode `\#12\catcode `\^12\catcode `\_12\catcode `\%12\relax}%
	\providecommand \@@startlink[1]{}%
	\providecommand \@@endlink[0]{}%
	\providecommand \url  [0]{\begingroup\@sanitize@url \@url }%
	\providecommand \@url [1]{\endgroup\@href {#1}{\urlprefix }}%
	\providecommand \urlprefix  [0]{URL }%
	\providecommand \Eprint [0]{\href }%
	\providecommand \doibase [0]{http://dx.doi.org/}%
	\providecommand \selectlanguage [0]{\@gobble}%
	\providecommand \bibinfo  [0]{\@secondoftwo}%
	\providecommand \bibfield  [0]{\@secondoftwo}%
	\providecommand \translation [1]{[#1]}%
	\providecommand \BibitemOpen [0]{}%
	\providecommand \bibitemStop [0]{}%
	\providecommand \bibitemNoStop [0]{.\EOS\space}%
	\providecommand \EOS [0]{\spacefactor3000\relax}%
	\providecommand \BibitemShut  [1]{\csname bibitem#1\endcsname}%
	\let\auto@bib@innerbib\@empty
	\bibitem [{\citenamefont {Kamihara}\ \emph {et~al.}(2008)\citenamefont
		{Kamihara}, \citenamefont {Watanabe}, \citenamefont {Hirano},\ and\
		\citenamefont {Hosono}}]{Kamihara2008}%
	\BibitemOpen
	\bibfield  {author} {\bibinfo {author} {\bibfnamefont {Y.}~\bibnamefont
			{Kamihara}}, \bibinfo {author} {\bibfnamefont {T.}~\bibnamefont {Watanabe}},
		\bibinfo {author} {\bibfnamefont {M.}~\bibnamefont {Hirano}}, \ and\ \bibinfo
		{author} {\bibfnamefont {H.}~\bibnamefont {Hosono}},\ }\href
	{http://dx.doi.org/10.1021/ja800073m} {\bibfield  {journal} {\bibinfo
			{journal} {Journal of the American Chemical Society}\ }\textbf {\bibinfo
			{volume} {130}},\ \bibinfo {pages} {3296} (\bibinfo {year}
		{2008})}\BibitemShut {NoStop}%
	\bibitem [{\citenamefont {Ren}\ \emph {et~al.}(2008)\citenamefont {Ren},
		\citenamefont {Che}, \citenamefont {Dong}, \citenamefont {Yang},
		\citenamefont {Lu}, \citenamefont {Yi}, \citenamefont {Shen}, \citenamefont
		{Li}, \citenamefont {Sun}, \citenamefont {Zhou},\ and\ \citenamefont
		{Zhao}}]{Ren2008}%
	\BibitemOpen
	\bibfield  {author} {\bibinfo {author} {\bibfnamefont {Z.~A.}\ \bibnamefont
			{Ren}}, \bibinfo {author} {\bibfnamefont {G.~C.}\ \bibnamefont {Che}},
		\bibinfo {author} {\bibfnamefont {X.~L.}\ \bibnamefont {Dong}}, \bibinfo
		{author} {\bibfnamefont {J.}~\bibnamefont {Yang}}, \bibinfo {author}
		{\bibfnamefont {W.}~\bibnamefont {Lu}}, \bibinfo {author} {\bibfnamefont
			{W.}~\bibnamefont {Yi}}, \bibinfo {author} {\bibfnamefont {X.~L.}\
			\bibnamefont {Shen}}, \bibinfo {author} {\bibfnamefont {Z.~C.}\ \bibnamefont
			{Li}}, \bibinfo {author} {\bibfnamefont {L.~L.}\ \bibnamefont {Sun}},
		\bibinfo {author} {\bibfnamefont {F.}~\bibnamefont {Zhou}}, \ and\ \bibinfo
		{author} {\bibfnamefont {Z.~X.}\ \bibnamefont {Zhao}},\ }\href
	{http://stacks.iop.org/0295-5075/83/i=1/a=17002} {\bibfield  {journal}
		{\bibinfo  {journal} {EPL (Europhysics Letters)}\ }\textbf {\bibinfo {volume}
			{83}},\ \bibinfo {pages} {17002} (\bibinfo {year} {2008})}\BibitemShut
	{NoStop}%
	\bibitem [{\citenamefont {Rotter}\ \emph {et~al.}(2008)\citenamefont {Rotter},
		\citenamefont {Tegel},\ and\ \citenamefont {Johrendt}}]{Rotter2008PRL}%
	\BibitemOpen
	\bibfield  {author} {\bibinfo {author} {\bibfnamefont {M.}~\bibnamefont
			{Rotter}}, \bibinfo {author} {\bibfnamefont {M.}~\bibnamefont {Tegel}}, \
		and\ \bibinfo {author} {\bibfnamefont {D.}~\bibnamefont {Johrendt}},\ }\href
	{\doibase 10.1103/PhysRevLett.101.107006} {\bibfield  {journal} {\bibinfo
			{journal} {Phys. Rev. Lett.}\ }\textbf {\bibinfo {volume} {101}},\ \bibinfo
		{pages} {107006} (\bibinfo {year} {2008})}\BibitemShut {NoStop}%
	\bibitem [{\citenamefont {Takahashi}\ \emph {et~al.}(2008)\citenamefont
		{Takahashi}, \citenamefont {Igawa}, \citenamefont {Arii}, \citenamefont
		{Kamihara}, \citenamefont {Hirano},\ and\ \citenamefont
		{Hosono}}]{Takahashi2008Nat}%
	\BibitemOpen
	\bibfield  {author} {\bibinfo {author} {\bibfnamefont {H.}~\bibnamefont
			{Takahashi}}, \bibinfo {author} {\bibfnamefont {K.}~\bibnamefont {Igawa}},
		\bibinfo {author} {\bibfnamefont {K.}~\bibnamefont {Arii}}, \bibinfo {author}
		{\bibfnamefont {Y.}~\bibnamefont {Kamihara}}, \bibinfo {author}
		{\bibfnamefont {M.}~\bibnamefont {Hirano}}, \ and\ \bibinfo {author}
		{\bibfnamefont {H.}~\bibnamefont {Hosono}},\ }\href
	{http://dx.doi.org/10.1038/nature06972} {\bibfield  {journal} {\bibinfo
			{journal} {Nature}\ }\textbf {\bibinfo {volume} {453}},\ \bibinfo {pages}
		{376} (\bibinfo {year} {2008})}\BibitemShut {NoStop}%
	\bibitem [{\citenamefont {Canfield}\ and\ \citenamefont
		{Bud'ko}(2010)}]{Canfield2010}%
	\BibitemOpen
	\bibfield  {author} {\bibinfo {author} {\bibfnamefont {P.~C.}\ \bibnamefont
			{Canfield}}\ and\ \bibinfo {author} {\bibfnamefont {S.~L.}\ \bibnamefont
			{Bud'ko}},\ }\href {\doibase 10.1146/annurev-conmatphys-070909-104041}
	{\bibfield  {journal} {\bibinfo  {journal} {Annu. Rev. Condens. Matter
				Phys.}\ }\textbf {\bibinfo {volume} {1}},\ \bibinfo {pages} {27} (\bibinfo
		{year} {2010})}\BibitemShut {NoStop}%
	\bibitem [{\citenamefont {Ni}\ and\ \citenamefont {Bud'ko}(2011)}]{Ni2011}%
	\BibitemOpen
	\bibfield  {author} {\bibinfo {author} {\bibfnamefont {N.}~\bibnamefont
			{Ni}}\ and\ \bibinfo {author} {\bibfnamefont {S.~L.}\ \bibnamefont
			{Bud'ko}},\ }\href {\doibase 10.1557/mrs.2011.178} {\bibfield  {journal}
		{\bibinfo  {journal} {MRS Bulletin}\ }\textbf {\bibinfo {volume} {36}},\
		\bibinfo {pages} {620–625} (\bibinfo {year} {2011})}\BibitemShut {NoStop}%
	\bibitem [{\citenamefont {Paglione}\ and\ \citenamefont
		{Greene}(2010)}]{Paglione2010}%
	\BibitemOpen
	\bibfield  {author} {\bibinfo {author} {\bibfnamefont {J.}~\bibnamefont
			{Paglione}}\ and\ \bibinfo {author} {\bibfnamefont {R.~L.}\ \bibnamefont
			{Greene}},\ }\href {http://dx.doi.org/10.1038/nphys1759} {\bibfield
		{journal} {\bibinfo  {journal} {Nat Phys}\ }\textbf {\bibinfo {volume} {6}},\
		\bibinfo {pages} {645} (\bibinfo {year} {2010})}\BibitemShut {NoStop}%
	\bibitem [{\citenamefont {Torikachvili}\ \emph
		{et~al.}(2008{\natexlab{a}})\citenamefont {Torikachvili}, \citenamefont
		{Bud'ko}, \citenamefont {Ni},\ and\ \citenamefont
		{Canfield}}]{Torikachvili2008PRB}%
	\BibitemOpen
	\bibfield  {author} {\bibinfo {author} {\bibfnamefont {M.~S.}\ \bibnamefont
			{Torikachvili}}, \bibinfo {author} {\bibfnamefont {S.~L.}\ \bibnamefont
			{Bud'ko}}, \bibinfo {author} {\bibfnamefont {N.}~\bibnamefont {Ni}}, \ and\
		\bibinfo {author} {\bibfnamefont {P.~C.}\ \bibnamefont {Canfield}},\ }\href
	{\doibase 10.1103/PhysRevB.78.104527} {\bibfield  {journal} {\bibinfo
			{journal} {Phys. Rev. B}\ }\textbf {\bibinfo {volume} {78}},\ \bibinfo
		{pages} {104527} (\bibinfo {year} {2008}{\natexlab{a}})}\BibitemShut
	{NoStop}%
	\bibitem [{\citenamefont {Alireza}\ \emph {et~al.}(2009)\citenamefont
		{Alireza}, \citenamefont {Ko}, \citenamefont {Gillett}, \citenamefont
		{Petrone}, \citenamefont {Cole}, \citenamefont {Lonzarich},\ and\
		\citenamefont {Sebastian}}]{Alireza2009}%
	\BibitemOpen
	\bibfield  {author} {\bibinfo {author} {\bibfnamefont {P.~L.}\ \bibnamefont
			{Alireza}}, \bibinfo {author} {\bibfnamefont {Y.~T.~C.}\ \bibnamefont {Ko}},
		\bibinfo {author} {\bibfnamefont {J.}~\bibnamefont {Gillett}}, \bibinfo
		{author} {\bibfnamefont {C.~M.}\ \bibnamefont {Petrone}}, \bibinfo {author}
		{\bibfnamefont {J.~M.}\ \bibnamefont {Cole}}, \bibinfo {author}
		{\bibfnamefont {G.~G.}\ \bibnamefont {Lonzarich}}, \ and\ \bibinfo {author}
		{\bibfnamefont {S.~E.}\ \bibnamefont {Sebastian}},\ }\href
	{http://stacks.iop.org/0953-8984/21/i=1/a=012208} {\bibfield  {journal}
		{\bibinfo  {journal} {Journal of Physics: Condensed Matter}\ }\textbf
		{\bibinfo {volume} {21}},\ \bibinfo {pages} {012208} (\bibinfo {year}
		{2009})}\BibitemShut {NoStop}%
	\bibitem [{\citenamefont {Kimber}\ \emph {et~al.}(2009)\citenamefont {Kimber},
		\citenamefont {Kreyssig}, \citenamefont {Zhang}, \citenamefont {Jeschke},
		\citenamefont {Valentí}, \citenamefont {Yokaichiya}, \citenamefont
		{Colombier}, \citenamefont {Yan}, \citenamefont {Hansen}, \citenamefont
		{Chatterji}, \citenamefont {McQueeney}, \citenamefont {Canfield},
		\citenamefont {Goldman},\ and\ \citenamefont {Argyriou}}]{Kimber2009}%
	\BibitemOpen
	\bibfield  {author} {\bibinfo {author} {\bibfnamefont {S.~A.~J.}\
			\bibnamefont {Kimber}}, \bibinfo {author} {\bibfnamefont {A.}~\bibnamefont
			{Kreyssig}}, \bibinfo {author} {\bibfnamefont {Y.-Z.}\ \bibnamefont {Zhang}},
		\bibinfo {author} {\bibfnamefont {H.~O.}\ \bibnamefont {Jeschke}}, \bibinfo
		{author} {\bibfnamefont {R.}~\bibnamefont {Valentí}}, \bibinfo {author}
		{\bibfnamefont {F.}~\bibnamefont {Yokaichiya}}, \bibinfo {author}
		{\bibfnamefont {E.}~\bibnamefont {Colombier}}, \bibinfo {author}
		{\bibfnamefont {J.}~\bibnamefont {Yan}}, \bibinfo {author} {\bibfnamefont
			{T.~C.}\ \bibnamefont {Hansen}}, \bibinfo {author} {\bibfnamefont
			{T.}~\bibnamefont {Chatterji}}, \bibinfo {author} {\bibfnamefont {R.~J.}\
			\bibnamefont {McQueeney}}, \bibinfo {author} {\bibfnamefont {P.~C.}\
			\bibnamefont {Canfield}}, \bibinfo {author} {\bibfnamefont {A.~I.}\
			\bibnamefont {Goldman}}, \ and\ \bibinfo {author} {\bibfnamefont {D.~N.}\
			\bibnamefont {Argyriou}},\ }\href {http://dx.doi.org/10.1038/nmat2443}
	{\bibfield  {journal} {\bibinfo  {journal} {Nature Materials}\ }\textbf
		{\bibinfo {volume} {8}},\ \bibinfo {pages} {471} (\bibinfo {year}
		{2009})}\BibitemShut {NoStop}%
	\bibitem [{\citenamefont {Colombier}\ \emph {et~al.}(2009)\citenamefont
		{Colombier}, \citenamefont {Bud'ko}, \citenamefont {Ni},\ and\ \citenamefont
		{Canfield}}]{Colombier2009}%
	\BibitemOpen
	\bibfield  {author} {\bibinfo {author} {\bibfnamefont {E.}~\bibnamefont
			{Colombier}}, \bibinfo {author} {\bibfnamefont {S.~L.}\ \bibnamefont
			{Bud'ko}}, \bibinfo {author} {\bibfnamefont {N.}~\bibnamefont {Ni}}, \ and\
		\bibinfo {author} {\bibfnamefont {P.~C.}\ \bibnamefont {Canfield}},\ }\href
	{\doibase 10.1103/PhysRevB.79.224518} {\bibfield  {journal} {\bibinfo
			{journal} {Phys. Rev. B}\ }\textbf {\bibinfo {volume} {79}},\ \bibinfo
		{pages} {224518} (\bibinfo {year} {2009})}\BibitemShut {NoStop}%
	\bibitem [{\citenamefont {Pratt}\ \emph {et~al.}(2009)\citenamefont {Pratt},
		\citenamefont {Tian}, \citenamefont {Kreyssig}, \citenamefont {Zarestky},
		\citenamefont {Nandi}, \citenamefont {Ni}, \citenamefont {Bud'ko},
		\citenamefont {Canfield}, \citenamefont {Goldman},\ and\ \citenamefont
		{McQueeney}}]{Pratt2009PRL}%
	\BibitemOpen
	\bibfield  {author} {\bibinfo {author} {\bibfnamefont {D.~K.}\ \bibnamefont
			{Pratt}}, \bibinfo {author} {\bibfnamefont {W.}~\bibnamefont {Tian}},
		\bibinfo {author} {\bibfnamefont {A.}~\bibnamefont {Kreyssig}}, \bibinfo
		{author} {\bibfnamefont {J.~L.}\ \bibnamefont {Zarestky}}, \bibinfo {author}
		{\bibfnamefont {S.}~\bibnamefont {Nandi}}, \bibinfo {author} {\bibfnamefont
			{N.}~\bibnamefont {Ni}}, \bibinfo {author} {\bibfnamefont {S.~L.}\
			\bibnamefont {Bud'ko}}, \bibinfo {author} {\bibfnamefont {P.~C.}\
			\bibnamefont {Canfield}}, \bibinfo {author} {\bibfnamefont {A.~I.}\
			\bibnamefont {Goldman}}, \ and\ \bibinfo {author} {\bibfnamefont {R.~J.}\
			\bibnamefont {McQueeney}},\ }\href {\doibase 10.1103/PhysRevLett.103.087001}
	{\bibfield  {journal} {\bibinfo  {journal} {Phys. Rev. Lett.}\ }\textbf
		{\bibinfo {volume} {103}},\ \bibinfo {pages} {087001} (\bibinfo {year}
		{2009})}\BibitemShut {NoStop}%
	\bibitem [{\citenamefont {Christianson}\ \emph {et~al.}(2009)\citenamefont
		{Christianson}, \citenamefont {Lumsden}, \citenamefont {Nagler},
		\citenamefont {MacDougall}, \citenamefont {McGuire}, \citenamefont {Sefat},
		\citenamefont {Jin}, \citenamefont {Sales},\ and\ \citenamefont
		{Mandrus}}]{Christianson2009PRL}%
	\BibitemOpen
	\bibfield  {author} {\bibinfo {author} {\bibfnamefont {A.~D.}\ \bibnamefont
			{Christianson}}, \bibinfo {author} {\bibfnamefont {M.~D.}\ \bibnamefont
			{Lumsden}}, \bibinfo {author} {\bibfnamefont {S.~E.}\ \bibnamefont {Nagler}},
		\bibinfo {author} {\bibfnamefont {G.~J.}\ \bibnamefont {MacDougall}},
		\bibinfo {author} {\bibfnamefont {M.~A.}\ \bibnamefont {McGuire}}, \bibinfo
		{author} {\bibfnamefont {A.~S.}\ \bibnamefont {Sefat}}, \bibinfo {author}
		{\bibfnamefont {R.}~\bibnamefont {Jin}}, \bibinfo {author} {\bibfnamefont
			{B.~C.}\ \bibnamefont {Sales}}, \ and\ \bibinfo {author} {\bibfnamefont
			{D.}~\bibnamefont {Mandrus}},\ }\href {\doibase
		10.1103/PhysRevLett.103.087002} {\bibfield  {journal} {\bibinfo  {journal}
			{Phys. Rev. Lett.}\ }\textbf {\bibinfo {volume} {103}},\ \bibinfo {pages}
		{087002} (\bibinfo {year} {2009})}\BibitemShut {NoStop}%
	\bibitem [{\citenamefont {Fernandes}\ \emph {et~al.}(2010)\citenamefont
		{Fernandes}, \citenamefont {Pratt}, \citenamefont {Tian}, \citenamefont
		{Zarestky}, \citenamefont {Kreyssig}, \citenamefont {Nandi}, \citenamefont
		{Kim}, \citenamefont {Thaler}, \citenamefont {Ni}, \citenamefont {Canfield},
		\citenamefont {McQueeney}, \citenamefont {Schmalian},\ and\ \citenamefont
		{Goldman}}]{Fernandes2010PRB}%
	\BibitemOpen
	\bibfield  {author} {\bibinfo {author} {\bibfnamefont {R.~M.}\ \bibnamefont
			{Fernandes}}, \bibinfo {author} {\bibfnamefont {D.~K.}\ \bibnamefont
			{Pratt}}, \bibinfo {author} {\bibfnamefont {W.}~\bibnamefont {Tian}},
		\bibinfo {author} {\bibfnamefont {J.}~\bibnamefont {Zarestky}}, \bibinfo
		{author} {\bibfnamefont {A.}~\bibnamefont {Kreyssig}}, \bibinfo {author}
		{\bibfnamefont {S.}~\bibnamefont {Nandi}}, \bibinfo {author} {\bibfnamefont
			{M.~G.}\ \bibnamefont {Kim}}, \bibinfo {author} {\bibfnamefont
			{A.}~\bibnamefont {Thaler}}, \bibinfo {author} {\bibfnamefont
			{N.}~\bibnamefont {Ni}}, \bibinfo {author} {\bibfnamefont {P.~C.}\
			\bibnamefont {Canfield}}, \bibinfo {author} {\bibfnamefont {R.~J.}\
			\bibnamefont {McQueeney}}, \bibinfo {author} {\bibfnamefont {J.}~\bibnamefont
			{Schmalian}}, \ and\ \bibinfo {author} {\bibfnamefont {A.~I.}\ \bibnamefont
			{Goldman}},\ }\href {\doibase 10.1103/PhysRevB.81.140501} {\bibfield
		{journal} {\bibinfo  {journal} {Phys. Rev. B}\ }\textbf {\bibinfo {volume}
			{81}},\ \bibinfo {pages} {140501} (\bibinfo {year} {2010})}\BibitemShut
	{NoStop}%
	\bibitem [{\citenamefont {Christianson}\ \emph {et~al.}(2008)\citenamefont
		{Christianson}, \citenamefont {Goremychkin}, \citenamefont {Osborn},
		\citenamefont {Rosenkranz}, \citenamefont {Lumsden}, \citenamefont
		{Malliakas}, \citenamefont {Todorov}, \citenamefont {Claus}, \citenamefont
		{Chung}, \citenamefont {Kanatzidis}, \citenamefont {Bewley},\ and\
		\citenamefont {Guidi}}]{Christianson2008Nat}%
	\BibitemOpen
	\bibfield  {author} {\bibinfo {author} {\bibfnamefont {A.~D.}\ \bibnamefont
			{Christianson}}, \bibinfo {author} {\bibfnamefont {E.~A.}\ \bibnamefont
			{Goremychkin}}, \bibinfo {author} {\bibfnamefont {R.}~\bibnamefont {Osborn}},
		\bibinfo {author} {\bibfnamefont {S.}~\bibnamefont {Rosenkranz}}, \bibinfo
		{author} {\bibfnamefont {M.~D.}\ \bibnamefont {Lumsden}}, \bibinfo {author}
		{\bibfnamefont {C.~D.}\ \bibnamefont {Malliakas}}, \bibinfo {author}
		{\bibfnamefont {I.~S.}\ \bibnamefont {Todorov}}, \bibinfo {author}
		{\bibfnamefont {H.}~\bibnamefont {Claus}}, \bibinfo {author} {\bibfnamefont
			{D.~Y.}\ \bibnamefont {Chung}}, \bibinfo {author} {\bibfnamefont {M.~G.}\
			\bibnamefont {Kanatzidis}}, \bibinfo {author} {\bibfnamefont {R.~I.}\
			\bibnamefont {Bewley}}, \ and\ \bibinfo {author} {\bibfnamefont
			{T.}~\bibnamefont {Guidi}},\ }\href {http://dx.doi.org/10.1038/nature07625}
	{\bibfield  {journal} {\bibinfo  {journal} {Nature}\ }\textbf {\bibinfo
			{volume} {456}},\ \bibinfo {pages} {930} (\bibinfo {year}
		{2008})}\BibitemShut {NoStop}%
	\bibitem [{\citenamefont {Yu}\ \emph {et~al.}(2009{\natexlab{a}})\citenamefont
		{Yu}, \citenamefont {Li}, \citenamefont {Motoyama},\ and\ \citenamefont
		{Greven}}]{Yu2009Nat}%
	\BibitemOpen
	\bibfield  {author} {\bibinfo {author} {\bibfnamefont {G.}~\bibnamefont
			{Yu}}, \bibinfo {author} {\bibfnamefont {Y.}~\bibnamefont {Li}}, \bibinfo
		{author} {\bibfnamefont {E.~M.}\ \bibnamefont {Motoyama}}, \ and\ \bibinfo
		{author} {\bibfnamefont {M.}~\bibnamefont {Greven}},\ }\href
	{http://dx.doi.org/10.1038/nphys1426} {\bibfield  {journal} {\bibinfo
			{journal} {Nature Physics}\ }\textbf {\bibinfo {volume} {5}},\ \bibinfo
		{pages} {873} (\bibinfo {year} {2009}{\natexlab{a}})}\BibitemShut {NoStop}%
	\bibitem [{\citenamefont {Iyo}\ \emph {et~al.}(2016)\citenamefont {Iyo},
		\citenamefont {Kawashima}, \citenamefont {Kinjo}, \citenamefont {Nishio},
		\citenamefont {Ishida}, \citenamefont {Fujihisa}, \citenamefont {Gotoh},
		\citenamefont {Kihou}, \citenamefont {Eisaki},\ and\ \citenamefont
		{Yoshida}}]{Iyo2016}%
	\BibitemOpen
	\bibfield  {author} {\bibinfo {author} {\bibfnamefont {A.}~\bibnamefont
			{Iyo}}, \bibinfo {author} {\bibfnamefont {K.}~\bibnamefont {Kawashima}},
		\bibinfo {author} {\bibfnamefont {T.}~\bibnamefont {Kinjo}}, \bibinfo
		{author} {\bibfnamefont {T.}~\bibnamefont {Nishio}}, \bibinfo {author}
		{\bibfnamefont {S.}~\bibnamefont {Ishida}}, \bibinfo {author} {\bibfnamefont
			{H.}~\bibnamefont {Fujihisa}}, \bibinfo {author} {\bibfnamefont
			{Y.}~\bibnamefont {Gotoh}}, \bibinfo {author} {\bibfnamefont
			{K.}~\bibnamefont {Kihou}}, \bibinfo {author} {\bibfnamefont
			{H.}~\bibnamefont {Eisaki}}, \ and\ \bibinfo {author} {\bibfnamefont
			{Y.}~\bibnamefont {Yoshida}},\ }\href {\doibase 10.1021/jacs.5b12571}
	{\bibfield  {journal} {\bibinfo  {journal} {Journal of the American Chemical
				Society}\ }\textbf {\bibinfo {volume} {138}},\ \bibinfo {pages} {3410}
		(\bibinfo {year} {2016})}\BibitemShut {NoStop}%
	\bibitem [{\citenamefont {Meier}\ \emph {et~al.}(2016)\citenamefont {Meier},
		\citenamefont {Kong}, \citenamefont {Kaluarachchi}, \citenamefont {Taufour},
		\citenamefont {Jo}, \citenamefont {Drachuck}, \citenamefont {B\"ohmer},
		\citenamefont {Saunders}, \citenamefont {Sapkota}, \citenamefont {Kreyssig},
		\citenamefont {Tanatar}, \citenamefont {Prozorov}, \citenamefont {Goldman},
		\citenamefont {Balakirev}, \citenamefont {Gurevich}, \citenamefont {Bud'ko},\
		and\ \citenamefont {Canfield}}]{Meier2016PRB}%
	\BibitemOpen
	\bibfield  {author} {\bibinfo {author} {\bibfnamefont {W.~R.}\ \bibnamefont
			{Meier}}, \bibinfo {author} {\bibfnamefont {T.}~\bibnamefont {Kong}},
		\bibinfo {author} {\bibfnamefont {U.~S.}\ \bibnamefont {Kaluarachchi}},
		\bibinfo {author} {\bibfnamefont {V.}~\bibnamefont {Taufour}}, \bibinfo
		{author} {\bibfnamefont {N.~H.}\ \bibnamefont {Jo}}, \bibinfo {author}
		{\bibfnamefont {G.}~\bibnamefont {Drachuck}}, \bibinfo {author}
		{\bibfnamefont {A.~E.}\ \bibnamefont {B\"ohmer}}, \bibinfo {author}
		{\bibfnamefont {S.~M.}\ \bibnamefont {Saunders}}, \bibinfo {author}
		{\bibfnamefont {A.}~\bibnamefont {Sapkota}}, \bibinfo {author} {\bibfnamefont
			{A.}~\bibnamefont {Kreyssig}}, \bibinfo {author} {\bibfnamefont {M.~A.}\
			\bibnamefont {Tanatar}}, \bibinfo {author} {\bibfnamefont {R.}~\bibnamefont
			{Prozorov}}, \bibinfo {author} {\bibfnamefont {A.~I.}\ \bibnamefont
			{Goldman}}, \bibinfo {author} {\bibfnamefont {F.~F.}\ \bibnamefont
			{Balakirev}}, \bibinfo {author} {\bibfnamefont {A.}~\bibnamefont {Gurevich}},
		\bibinfo {author} {\bibfnamefont {S.~L.}\ \bibnamefont {Bud'ko}}, \ and\
		\bibinfo {author} {\bibfnamefont {P.~C.}\ \bibnamefont {Canfield}},\ }\href
	{\doibase 10.1103/PhysRevB.94.064501} {\bibfield  {journal} {\bibinfo
			{journal} {Phys. Rev. B}\ }\textbf {\bibinfo {volume} {94}},\ \bibinfo
		{pages} {064501} (\bibinfo {year} {2016})}\BibitemShut {NoStop}%
	\bibitem [{\citenamefont {Meier}\ \emph {et~al.}(2017)\citenamefont {Meier},
		\citenamefont {Kong}, \citenamefont {Bud'ko},\ and\ \citenamefont
		{Canfield}}]{Meier2017PRM}%
	\BibitemOpen
	\bibfield  {author} {\bibinfo {author} {\bibfnamefont {W.~R.}\ \bibnamefont
			{Meier}}, \bibinfo {author} {\bibfnamefont {T.}~\bibnamefont {Kong}},
		\bibinfo {author} {\bibfnamefont {S.~L.}\ \bibnamefont {Bud'ko}}, \ and\
		\bibinfo {author} {\bibfnamefont {P.~C.}\ \bibnamefont {Canfield}},\ }\href
	{\doibase 10.1103/PhysRevMaterials.1.013401} {\bibfield  {journal} {\bibinfo
			{journal} {Phys. Rev. Materials}\ }\textbf {\bibinfo {volume} {1}},\ \bibinfo
		{pages} {013401} (\bibinfo {year} {2017})}\BibitemShut {NoStop}%
	\bibitem [{\citenamefont {Kaluarachchi}\ \emph {et~al.}(2017)\citenamefont
		{Kaluarachchi}, \citenamefont {Taufour}, \citenamefont {Sapkota},
		\citenamefont {Borisov}, \citenamefont {Kong}, \citenamefont {Meier},
		\citenamefont {Kothapalli}, \citenamefont {Ueland}, \citenamefont {Kreyssig},
		\citenamefont {Valent\'{\i}}, \citenamefont {McQueeney}, \citenamefont
		{Goldman}, \citenamefont {Bud'ko},\ and\ \citenamefont
		{Canfield}}]{Kaluarachchi2017PRB}%
	\BibitemOpen
	\bibfield  {author} {\bibinfo {author} {\bibfnamefont {U.~S.}\ \bibnamefont
			{Kaluarachchi}}, \bibinfo {author} {\bibfnamefont {V.}~\bibnamefont
			{Taufour}}, \bibinfo {author} {\bibfnamefont {A.}~\bibnamefont {Sapkota}},
		\bibinfo {author} {\bibfnamefont {V.}~\bibnamefont {Borisov}}, \bibinfo
		{author} {\bibfnamefont {T.}~\bibnamefont {Kong}}, \bibinfo {author}
		{\bibfnamefont {W.~R.}\ \bibnamefont {Meier}}, \bibinfo {author}
		{\bibfnamefont {K.}~\bibnamefont {Kothapalli}}, \bibinfo {author}
		{\bibfnamefont {B.~G.}\ \bibnamefont {Ueland}}, \bibinfo {author}
		{\bibfnamefont {A.}~\bibnamefont {Kreyssig}}, \bibinfo {author}
		{\bibfnamefont {R.}~\bibnamefont {Valent\'{\i}}}, \bibinfo {author}
		{\bibfnamefont {R.~J.}\ \bibnamefont {McQueeney}}, \bibinfo {author}
		{\bibfnamefont {A.~I.}\ \bibnamefont {Goldman}}, \bibinfo {author}
		{\bibfnamefont {S.~L.}\ \bibnamefont {Bud'ko}}, \ and\ \bibinfo {author}
		{\bibfnamefont {P.~C.}\ \bibnamefont {Canfield}},\ }\href {\doibase
		10.1103/PhysRevB.96.140501} {\bibfield  {journal} {\bibinfo  {journal} {Phys.
				Rev. B}\ }\textbf {\bibinfo {volume} {96}},\ \bibinfo {pages} {140501}
		(\bibinfo {year} {2017})}\BibitemShut {NoStop}%
	\bibitem [{\citenamefont {Torikachvili}\ \emph
		{et~al.}(2008{\natexlab{b}})\citenamefont {Torikachvili}, \citenamefont
		{Bud'ko}, \citenamefont {Ni},\ and\ \citenamefont
		{Canfield}}]{Torikachvili2008PRL}%
	\BibitemOpen
	\bibfield  {author} {\bibinfo {author} {\bibfnamefont {M.~S.}\ \bibnamefont
			{Torikachvili}}, \bibinfo {author} {\bibfnamefont {S.~L.}\ \bibnamefont
			{Bud'ko}}, \bibinfo {author} {\bibfnamefont {N.}~\bibnamefont {Ni}}, \ and\
		\bibinfo {author} {\bibfnamefont {P.~C.}\ \bibnamefont {Canfield}},\ }\href
	{\doibase 10.1103/PhysRevLett.101.057006} {\bibfield  {journal} {\bibinfo
			{journal} {Phys. Rev. Lett.}\ }\textbf {\bibinfo {volume} {101}},\ \bibinfo
		{pages} {057006} (\bibinfo {year} {2008}{\natexlab{b}})}\BibitemShut
	{NoStop}%
	\bibitem [{\citenamefont {Yu}\ \emph {et~al.}(2009{\natexlab{b}})\citenamefont
		{Yu}, \citenamefont {Aczel}, \citenamefont {Williams}, \citenamefont
		{Bud'ko}, \citenamefont {Ni}, \citenamefont {Canfield},\ and\ \citenamefont
		{Luke}}]{Yu2009PRB}%
	\BibitemOpen
	\bibfield  {author} {\bibinfo {author} {\bibfnamefont {W.}~\bibnamefont
			{Yu}}, \bibinfo {author} {\bibfnamefont {A.~A.}\ \bibnamefont {Aczel}},
		\bibinfo {author} {\bibfnamefont {T.~J.}\ \bibnamefont {Williams}}, \bibinfo
		{author} {\bibfnamefont {S.~L.}\ \bibnamefont {Bud'ko}}, \bibinfo {author}
		{\bibfnamefont {N.}~\bibnamefont {Ni}}, \bibinfo {author} {\bibfnamefont
			{P.~C.}\ \bibnamefont {Canfield}}, \ and\ \bibinfo {author} {\bibfnamefont
			{G.~M.}\ \bibnamefont {Luke}},\ }\href {\doibase 10.1103/PhysRevB.79.020511}
	{\bibfield  {journal} {\bibinfo  {journal} {Phys. Rev. B}\ }\textbf {\bibinfo
			{volume} {79}},\ \bibinfo {pages} {020511} (\bibinfo {year}
		{2009}{\natexlab{b}})}\BibitemShut {NoStop}%
	\bibitem [{\citenamefont {Kreyssig}\ \emph {et~al.}(2008)\citenamefont
		{Kreyssig}, \citenamefont {Green}, \citenamefont {Lee}, \citenamefont
		{Samolyuk}, \citenamefont {Zajdel}, \citenamefont {Lynn}, \citenamefont
		{Bud'ko}, \citenamefont {Torikachvili}, \citenamefont {Ni}, \citenamefont
		{Nandi}, \citenamefont {Le\~ao}, \citenamefont {Poulton}, \citenamefont
		{Argyriou}, \citenamefont {Harmon}, \citenamefont {McQueeney}, \citenamefont
		{Canfield},\ and\ \citenamefont {Goldman}}]{Kreyssig2008PRB}%
	\BibitemOpen
	\bibfield  {author} {\bibinfo {author} {\bibfnamefont {A.}~\bibnamefont
			{Kreyssig}}, \bibinfo {author} {\bibfnamefont {M.~A.}\ \bibnamefont {Green}},
		\bibinfo {author} {\bibfnamefont {Y.}~\bibnamefont {Lee}}, \bibinfo {author}
		{\bibfnamefont {G.~D.}\ \bibnamefont {Samolyuk}}, \bibinfo {author}
		{\bibfnamefont {P.}~\bibnamefont {Zajdel}}, \bibinfo {author} {\bibfnamefont
			{J.~W.}\ \bibnamefont {Lynn}}, \bibinfo {author} {\bibfnamefont {S.~L.}\
			\bibnamefont {Bud'ko}}, \bibinfo {author} {\bibfnamefont {M.~S.}\
			\bibnamefont {Torikachvili}}, \bibinfo {author} {\bibfnamefont
			{N.}~\bibnamefont {Ni}}, \bibinfo {author} {\bibfnamefont {S.}~\bibnamefont
			{Nandi}}, \bibinfo {author} {\bibfnamefont {J.~B.}\ \bibnamefont {Le\~ao}},
		\bibinfo {author} {\bibfnamefont {S.~J.}\ \bibnamefont {Poulton}}, \bibinfo
		{author} {\bibfnamefont {D.~N.}\ \bibnamefont {Argyriou}}, \bibinfo {author}
		{\bibfnamefont {B.~N.}\ \bibnamefont {Harmon}}, \bibinfo {author}
		{\bibfnamefont {R.~J.}\ \bibnamefont {McQueeney}}, \bibinfo {author}
		{\bibfnamefont {P.~C.}\ \bibnamefont {Canfield}}, \ and\ \bibinfo {author}
		{\bibfnamefont {A.~I.}\ \bibnamefont {Goldman}},\ }\href {\doibase
		10.1103/PhysRevB.78.184517} {\bibfield  {journal} {\bibinfo  {journal} {Phys.
				Rev. B}\ }\textbf {\bibinfo {volume} {78}},\ \bibinfo {pages} {184517}
		(\bibinfo {year} {2008})}\BibitemShut {NoStop}%
	\bibitem [{\citenamefont {Meier}\ \emph {et~al.}(2018)\citenamefont {Meier},
		\citenamefont {Ding}, \citenamefont {Kreyssig}, \citenamefont {Bud‘ko},
		\citenamefont {Sapkota}, \citenamefont {Kothapalli}, \citenamefont {Borisov},
		\citenamefont {Valent\'i}, \citenamefont {Batista}, \citenamefont {Orth},
		\citenamefont {Fernandes}, \citenamefont {Goldman}, \citenamefont {Furukawa},
		\citenamefont {B\"ohmer},\ and\ \citenamefont {Canfield}}]{Meier2018}%
	\BibitemOpen
	\bibfield  {author} {\bibinfo {author} {\bibfnamefont {W.~R.}\ \bibnamefont
			{Meier}}, \bibinfo {author} {\bibfnamefont {Q.-P.}\ \bibnamefont {Ding}},
		\bibinfo {author} {\bibfnamefont {A.}~\bibnamefont {Kreyssig}}, \bibinfo
		{author} {\bibfnamefont {S.~L.}\ \bibnamefont {Bud‘ko}}, \bibinfo {author}
		{\bibfnamefont {A.}~\bibnamefont {Sapkota}}, \bibinfo {author} {\bibfnamefont
			{K.}~\bibnamefont {Kothapalli}}, \bibinfo {author} {\bibfnamefont
			{V.}~\bibnamefont {Borisov}}, \bibinfo {author} {\bibfnamefont
			{R.}~\bibnamefont {Valent\'i}}, \bibinfo {author} {\bibfnamefont {C.~D.}\
			\bibnamefont {Batista}}, \bibinfo {author} {\bibfnamefont {P.~P.}\
			\bibnamefont {Orth}}, \bibinfo {author} {\bibfnamefont {R.~M.}\ \bibnamefont
			{Fernandes}}, \bibinfo {author} {\bibfnamefont {A.~I.}\ \bibnamefont
			{Goldman}}, \bibinfo {author} {\bibfnamefont {Y.}~\bibnamefont {Furukawa}},
		\bibinfo {author} {\bibfnamefont {A.~E.}\ \bibnamefont {B\"ohmer}}, \ and\
		\bibinfo {author} {\bibfnamefont {P.~C.}\ \bibnamefont {Canfield}},\ }\href
	{https://doi.org/10.1038/s41535-017-0076-x} {\bibfield  {journal} {\bibinfo
			{journal} {npj Quantum Materials}\ }\textbf {\bibinfo {volume} {3}},\
		\bibinfo {pages} {5} (\bibinfo {year} {2018})}\BibitemShut {NoStop}%
	\bibitem [{\citenamefont {Fernandes}\ \emph {et~al.}(2016)\citenamefont
		{Fernandes}, \citenamefont {Kivelson},\ and\ \citenamefont
		{Berg}}]{Fernandes2016PRB}%
	\BibitemOpen
	\bibfield  {author} {\bibinfo {author} {\bibfnamefont {R.~M.}\ \bibnamefont
			{Fernandes}}, \bibinfo {author} {\bibfnamefont {S.~A.}\ \bibnamefont
			{Kivelson}}, \ and\ \bibinfo {author} {\bibfnamefont {E.}~\bibnamefont
			{Berg}},\ }\href {\doibase 10.1103/PhysRevB.93.014511} {\bibfield  {journal}
		{\bibinfo  {journal} {Phys. Rev. B}\ }\textbf {\bibinfo {volume} {93}},\
		\bibinfo {pages} {014511} (\bibinfo {year} {2016})}\BibitemShut {NoStop}%
	\bibitem [{\citenamefont {Cvetkovic}\ and\ \citenamefont
		{Vafek}(2013)}]{Cvetkovic2013PRB}%
	\BibitemOpen
	\bibfield  {author} {\bibinfo {author} {\bibfnamefont {V.}~\bibnamefont
			{Cvetkovic}}\ and\ \bibinfo {author} {\bibfnamefont {O.}~\bibnamefont
			{Vafek}},\ }\href {\doibase 10.1103/PhysRevB.88.134510} {\bibfield  {journal}
		{\bibinfo  {journal} {Phys. Rev. B}\ }\textbf {\bibinfo {volume} {88}},\
		\bibinfo {pages} {134510} (\bibinfo {year} {2013})}\BibitemShut {NoStop}%
	\bibitem [{\citenamefont {O'Halloran}\ \emph {et~al.}(2017)\citenamefont
		{O'Halloran}, \citenamefont {Agterberg}, \citenamefont {Chen},\ and\
		\citenamefont {Weinert}}]{Halloran2017PRB}%
	\BibitemOpen
	\bibfield  {author} {\bibinfo {author} {\bibfnamefont {J.}~\bibnamefont
			{O'Halloran}}, \bibinfo {author} {\bibfnamefont {D.~F.}\ \bibnamefont
			{Agterberg}}, \bibinfo {author} {\bibfnamefont {M.~X.}\ \bibnamefont {Chen}},
		\ and\ \bibinfo {author} {\bibfnamefont {M.}~\bibnamefont {Weinert}},\ }\href
	{\doibase 10.1103/PhysRevB.95.075104} {\bibfield  {journal} {\bibinfo
			{journal} {Phys. Rev. B}\ }\textbf {\bibinfo {volume} {95}},\ \bibinfo
		{pages} {075104} (\bibinfo {year} {2017})}\BibitemShut {NoStop}%
	\bibitem [{\citenamefont {Colombier}\ \emph {et~al.}(2010)\citenamefont
		{Colombier}, \citenamefont {Torikachvili}, \citenamefont {Ni}, \citenamefont
		{Thaler}, \citenamefont {Bud'ko},\ and\ \citenamefont
		{Canfield}}]{Colombier2010}%
	\BibitemOpen
	\bibfield  {author} {\bibinfo {author} {\bibfnamefont {E.}~\bibnamefont
			{Colombier}}, \bibinfo {author} {\bibfnamefont {M.~S.}\ \bibnamefont
			{Torikachvili}}, \bibinfo {author} {\bibfnamefont {N.}~\bibnamefont {Ni}},
		\bibinfo {author} {\bibfnamefont {A.}~\bibnamefont {Thaler}}, \bibinfo
		{author} {\bibfnamefont {S.~L.}\ \bibnamefont {Bud'ko}}, \ and\ \bibinfo
		{author} {\bibfnamefont {P.~C.}\ \bibnamefont {Canfield}},\ }\href
	{http://stacks.iop.org/0953-2048/23/i=5/a=054003} {\bibfield  {journal}
		{\bibinfo  {journal} {Superconductor Science and Technology}\ }\textbf
		{\bibinfo {volume} {23}},\ \bibinfo {pages} {054003} (\bibinfo {year}
		{2010})}\BibitemShut {NoStop}%
	\bibitem [{\citenamefont {Bud'ko}\ \emph {et~al.}(1984)\citenamefont {Bud'ko},
		\citenamefont {Voronovskii}, \citenamefont {Gapotchenko},\ and\ \citenamefont
		{ltskevich}}]{Budko1984}%
	\BibitemOpen
	\bibfield  {author} {\bibinfo {author} {\bibfnamefont {S.~L.}\ \bibnamefont
			{Bud'ko}}, \bibinfo {author} {\bibfnamefont {A.~N.}\ \bibnamefont
			{Voronovskii}}, \bibinfo {author} {\bibfnamefont {A.~G.}\ \bibnamefont
			{Gapotchenko}}, \ and\ \bibinfo {author} {\bibfnamefont {E.~S.}\ \bibnamefont
			{ltskevich}},\ }\href
	{http://www.jetp.ac.ru/cgi-bin/e/index/e/59/2/p454?a=list} {\bibfield
		{journal} {\bibinfo  {journal} {Zh. Eksp. Teor. Fiz. 86, 778}\ } (\bibinfo
		{year} {1984})}\BibitemShut {NoStop}%
	\bibitem [{\citenamefont {Colombier}\ and\ \citenamefont
		{Braithwaite}(2007)}]{Colombier2007}%
	\BibitemOpen
	\bibfield  {author} {\bibinfo {author} {\bibfnamefont {E.}~\bibnamefont
			{Colombier}}\ and\ \bibinfo {author} {\bibfnamefont {D.}~\bibnamefont
			{Braithwaite}},\ }\href {http://aip.scitation.org/doi/full/10.1063/1.2778629}
	{\bibfield  {journal} {\bibinfo  {journal} {Review of Scientific Instruments
				78, 093903}\ } (\bibinfo {year} {2007})}\BibitemShut {NoStop}%
	\bibitem [{\citenamefont {Bireckoven}\ and\ \citenamefont
		{Wittig}(1988)}]{Bireckoven1988}%
	\BibitemOpen
	\bibfield  {author} {\bibinfo {author} {\bibfnamefont {B.}~\bibnamefont
			{Bireckoven}}\ and\ \bibinfo {author} {\bibfnamefont {J.}~\bibnamefont
			{Wittig}},\ }\href {http://stacks.iop.org/0022-3735/21/i=9/a=004} {\bibfield
		{journal} {\bibinfo  {journal} {Journal of Physics E: Scientific
				Instruments}\ }\textbf {\bibinfo {volume} {21}},\ \bibinfo {pages} {841}
		(\bibinfo {year} {1988})}\BibitemShut {NoStop}%
	\bibitem [{\citenamefont {Kim}\ \emph {et~al.}(2011)\citenamefont {Kim},
		\citenamefont {Torikachvili}, \citenamefont {Colombier}, \citenamefont
		{Thaler}, \citenamefont {Bud'ko},\ and\ \citenamefont {Canfield}}]{Kim2011}%
	\BibitemOpen
	\bibfield  {author} {\bibinfo {author} {\bibfnamefont {S.~K.}\ \bibnamefont
			{Kim}}, \bibinfo {author} {\bibfnamefont {M.~S.}\ \bibnamefont
			{Torikachvili}}, \bibinfo {author} {\bibfnamefont {E.}~\bibnamefont
			{Colombier}}, \bibinfo {author} {\bibfnamefont {A.}~\bibnamefont {Thaler}},
		\bibinfo {author} {\bibfnamefont {S.~L.}\ \bibnamefont {Bud'ko}}, \ and\
		\bibinfo {author} {\bibfnamefont {P.~C.}\ \bibnamefont {Canfield}},\ }\href
	{\doibase 10.1103/PhysRevB.84.134525} {\bibfield  {journal} {\bibinfo
			{journal} {Phys. Rev. B}\ }\textbf {\bibinfo {volume} {84}},\ \bibinfo
		{pages} {134525} (\bibinfo {year} {2011})}\BibitemShut {NoStop}%
	\bibitem [{\citenamefont {Torikachvili}\ \emph {et~al.}(2015)\citenamefont
		{Torikachvili}, \citenamefont {Kim}, \citenamefont {Colombier}, \citenamefont
		{Bud'ko},\ and\ \citenamefont {Canfield}}]{Torikachvili2015}%
	\BibitemOpen
	\bibfield  {author} {\bibinfo {author} {\bibfnamefont {M.~S.}\ \bibnamefont
			{Torikachvili}}, \bibinfo {author} {\bibfnamefont {S.~K.}\ \bibnamefont
			{Kim}}, \bibinfo {author} {\bibfnamefont {E.}~\bibnamefont {Colombier}},
		\bibinfo {author} {\bibfnamefont {S.~L.}\ \bibnamefont {Bud'ko}}, \ and\
		\bibinfo {author} {\bibfnamefont {P.~C.}\ \bibnamefont {Canfield}},\ }\href
	{http://aip.scitation.org/doi/abs/10.1063/1.4937478} {\bibfield  {journal}
		{\bibinfo  {journal} {Rev. Sci. Instrum.}\ }\textbf {\bibinfo {volume}
			{86}},\ \bibinfo {pages} {123904} (\bibinfo {year} {2015})}\BibitemShut
	{NoStop}%
	\bibitem [{\citenamefont {Kogan}\ and\ \citenamefont
		{Prozorov}(2012)}]{Kogan2012}%
	\BibitemOpen
	\bibfield  {author} {\bibinfo {author} {\bibfnamefont {V.~G.}\ \bibnamefont
			{Kogan}}\ and\ \bibinfo {author} {\bibfnamefont {R.}~\bibnamefont
			{Prozorov}},\ }\href {http://stacks.iop.org/0034-4885/75/i=11/a=114502}
	{\bibfield  {journal} {\bibinfo  {journal} {Rep. Prog. Phys.}\ }\textbf
		{\bibinfo {volume} {75}},\ \bibinfo {pages} {114502} (\bibinfo {year}
		{2012})}\BibitemShut {NoStop}%
	\bibitem [{\citenamefont {Kaluarachchi}\ \emph {et~al.}(2016)\citenamefont
		{Kaluarachchi}, \citenamefont {Taufour}, \citenamefont {B\"ohmer},
		\citenamefont {Tanatar}, \citenamefont {Bud'ko}, \citenamefont {Kogan},
		\citenamefont {Prozorov},\ and\ \citenamefont {Canfield}}]{Kaluarachchi2016}%
	\BibitemOpen
	\bibfield  {author} {\bibinfo {author} {\bibfnamefont {U.~S.}\ \bibnamefont
			{Kaluarachchi}}, \bibinfo {author} {\bibfnamefont {V.}~\bibnamefont
			{Taufour}}, \bibinfo {author} {\bibfnamefont {A.~E.}\ \bibnamefont
			{B\"ohmer}}, \bibinfo {author} {\bibfnamefont {M.~A.}\ \bibnamefont
			{Tanatar}}, \bibinfo {author} {\bibfnamefont {S.~L.}\ \bibnamefont {Bud'ko}},
		\bibinfo {author} {\bibfnamefont {V.~G.}\ \bibnamefont {Kogan}}, \bibinfo
		{author} {\bibfnamefont {R.}~\bibnamefont {Prozorov}}, \ and\ \bibinfo
		{author} {\bibfnamefont {P.~C.}\ \bibnamefont {Canfield}},\ }\href {\doibase
		10.1103/PhysRevB.93.064503} {\bibfield  {journal} {\bibinfo  {journal} {Phys.
				Rev. B}\ }\textbf {\bibinfo {volume} {93}},\ \bibinfo {pages} {064503}
		(\bibinfo {year} {2016})}\BibitemShut {NoStop}%
	\bibitem [{\citenamefont {Xiang}\ \emph {et~al.}(2017)\citenamefont {Xiang},
		\citenamefont {Kaluarachchi}, \citenamefont {B\"ohmer}, \citenamefont
		{Taufour}, \citenamefont {Tanatar}, \citenamefont {Prozorov}, \citenamefont
		{Bud'ko},\ and\ \citenamefont {Canfield}}]{Xiang2017PRB}%
	\BibitemOpen
	\bibfield  {author} {\bibinfo {author} {\bibfnamefont {L.}~\bibnamefont
			{Xiang}}, \bibinfo {author} {\bibfnamefont {U.~S.}\ \bibnamefont
			{Kaluarachchi}}, \bibinfo {author} {\bibfnamefont {A.~E.}\ \bibnamefont
			{B\"ohmer}}, \bibinfo {author} {\bibfnamefont {V.}~\bibnamefont {Taufour}},
		\bibinfo {author} {\bibfnamefont {M.~A.}\ \bibnamefont {Tanatar}}, \bibinfo
		{author} {\bibfnamefont {R.}~\bibnamefont {Prozorov}}, \bibinfo {author}
		{\bibfnamefont {S.~L.}\ \bibnamefont {Bud'ko}}, \ and\ \bibinfo {author}
		{\bibfnamefont {P.~C.}\ \bibnamefont {Canfield}},\ }\href {\doibase
		10.1103/PhysRevB.96.024511} {\bibfield  {journal} {\bibinfo  {journal} {Phys.
				Rev. B}\ }\textbf {\bibinfo {volume} {96}},\ \bibinfo {pages} {024511}
		(\bibinfo {year} {2017})}\BibitemShut {NoStop}%
	\bibitem [{\citenamefont {Mou}\ \emph {et~al.}(2016)\citenamefont {Mou},
		\citenamefont {Kong}, \citenamefont {Meier}, \citenamefont {Lochner},
		\citenamefont {Wang}, \citenamefont {Lin}, \citenamefont {Wu}, \citenamefont
		{Bud'ko}, \citenamefont {Eremin}, \citenamefont {Johnson}, \citenamefont
		{Canfield},\ and\ \citenamefont {Kaminski}}]{Mou2016PRL}%
	\BibitemOpen
	\bibfield  {author} {\bibinfo {author} {\bibfnamefont {D.}~\bibnamefont
			{Mou}}, \bibinfo {author} {\bibfnamefont {T.}~\bibnamefont {Kong}}, \bibinfo
		{author} {\bibfnamefont {W.~R.}\ \bibnamefont {Meier}}, \bibinfo {author}
		{\bibfnamefont {F.}~\bibnamefont {Lochner}}, \bibinfo {author} {\bibfnamefont
			{L.-L.}\ \bibnamefont {Wang}}, \bibinfo {author} {\bibfnamefont
			{Q.}~\bibnamefont {Lin}}, \bibinfo {author} {\bibfnamefont {Y.}~\bibnamefont
			{Wu}}, \bibinfo {author} {\bibfnamefont {S.~L.}\ \bibnamefont {Bud'ko}},
		\bibinfo {author} {\bibfnamefont {I.}~\bibnamefont {Eremin}}, \bibinfo
		{author} {\bibfnamefont {D.~D.}\ \bibnamefont {Johnson}}, \bibinfo {author}
		{\bibfnamefont {P.~C.}\ \bibnamefont {Canfield}}, \ and\ \bibinfo {author}
		{\bibfnamefont {A.}~\bibnamefont {Kaminski}},\ }\href {\doibase
		10.1103/PhysRevLett.117.277001} {\bibfield  {journal} {\bibinfo  {journal}
			{Phys. Rev. Lett.}\ }\textbf {\bibinfo {volume} {117}},\ \bibinfo {pages}
		{277001} (\bibinfo {year} {2016})}\BibitemShut {NoStop}%
	\bibitem [{\citenamefont {Taufour}\ \emph {et~al.}(2014)\citenamefont
		{Taufour}, \citenamefont {Foroozani}, \citenamefont {Tanatar}, \citenamefont
		{Lim}, \citenamefont {Kaluarachchi}, \citenamefont {Kim}, \citenamefont
		{Liu}, \citenamefont {Lograsso}, \citenamefont {Kogan}, \citenamefont
		{Prozorov}, \citenamefont {Bud'ko}, \citenamefont {Schilling},\ and\
		\citenamefont {Canfield}}]{Taufour2014}%
	\BibitemOpen
	\bibfield  {author} {\bibinfo {author} {\bibfnamefont {V.}~\bibnamefont
			{Taufour}}, \bibinfo {author} {\bibfnamefont {N.}~\bibnamefont {Foroozani}},
		\bibinfo {author} {\bibfnamefont {M.~A.}\ \bibnamefont {Tanatar}}, \bibinfo
		{author} {\bibfnamefont {J.}~\bibnamefont {Lim}}, \bibinfo {author}
		{\bibfnamefont {U.}~\bibnamefont {Kaluarachchi}}, \bibinfo {author}
		{\bibfnamefont {S.~K.}\ \bibnamefont {Kim}}, \bibinfo {author} {\bibfnamefont
			{Y.}~\bibnamefont {Liu}}, \bibinfo {author} {\bibfnamefont {T.~A.}\
			\bibnamefont {Lograsso}}, \bibinfo {author} {\bibfnamefont {V.~G.}\
			\bibnamefont {Kogan}}, \bibinfo {author} {\bibfnamefont {R.}~\bibnamefont
			{Prozorov}}, \bibinfo {author} {\bibfnamefont {S.~L.}\ \bibnamefont
			{Bud'ko}}, \bibinfo {author} {\bibfnamefont {J.~S.}\ \bibnamefont
			{Schilling}}, \ and\ \bibinfo {author} {\bibfnamefont {P.~C.}\ \bibnamefont
			{Canfield}},\ }\href {\doibase 10.1103/PhysRevB.89.220509} {\bibfield
		{journal} {\bibinfo  {journal} {Phys. Rev. B}\ }\textbf {\bibinfo {volume}
			{89}},\ \bibinfo {pages} {220509} (\bibinfo {year} {2014})}\BibitemShut
	{NoStop}%
	\bibitem [{\citenamefont {Jiang}\ \emph {et~al.}(2009)\citenamefont {Jiang},
		\citenamefont {Xing}, \citenamefont {Xuan}, \citenamefont {Wang},
		\citenamefont {Ren}, \citenamefont {Feng}, \citenamefont {Dai}, \citenamefont
		{Xu},\ and\ \citenamefont {Cao}}]{Jiang2009}%
	\BibitemOpen
	\bibfield  {author} {\bibinfo {author} {\bibfnamefont {S.}~\bibnamefont
			{Jiang}}, \bibinfo {author} {\bibfnamefont {H.}~\bibnamefont {Xing}},
		\bibinfo {author} {\bibfnamefont {G.}~\bibnamefont {Xuan}}, \bibinfo {author}
		{\bibfnamefont {C.}~\bibnamefont {Wang}}, \bibinfo {author} {\bibfnamefont
			{Z.}~\bibnamefont {Ren}}, \bibinfo {author} {\bibfnamefont {C.}~\bibnamefont
			{Feng}}, \bibinfo {author} {\bibfnamefont {J.}~\bibnamefont {Dai}}, \bibinfo
		{author} {\bibfnamefont {Z.}~\bibnamefont {Xu}}, \ and\ \bibinfo {author}
		{\bibfnamefont {G.}~\bibnamefont {Cao}},\ }\href
	{http://stacks.iop.org/0953-8984/21/i=38/a=382203} {\bibfield  {journal}
		{\bibinfo  {journal} {J. Phys. Condens. Matter}\ }\textbf {\bibinfo {volume}
			{21}},\ \bibinfo {pages} {382203} (\bibinfo {year} {2009})}\BibitemShut
	{NoStop}%
	\bibitem [{\citenamefont {Dai}\ \emph {et~al.}(2009)\citenamefont {Dai},
		\citenamefont {Si}, \citenamefont {Zhu},\ and\ \citenamefont
		{Abrahams}}]{Dai2009}%
	\BibitemOpen
	\bibfield  {author} {\bibinfo {author} {\bibfnamefont {J.}~\bibnamefont
			{Dai}}, \bibinfo {author} {\bibfnamefont {Q.}~\bibnamefont {Si}}, \bibinfo
		{author} {\bibfnamefont {J.-X.}\ \bibnamefont {Zhu}}, \ and\ \bibinfo
		{author} {\bibfnamefont {E.}~\bibnamefont {Abrahams}},\ }\href {\doibase
		10.1073/pnas.0900886106} {\bibfield  {journal} {\bibinfo  {journal}
			{Proceedings of the National Academy of Sciences}\ }\textbf {\bibinfo
			{volume} {106}},\ \bibinfo {pages} {4118} (\bibinfo {year}
		{2009})}\BibitemShut {NoStop}%
	\bibitem [{\citenamefont {Gooch}\ \emph {et~al.}(2009)\citenamefont {Gooch},
		\citenamefont {Lv}, \citenamefont {Lorenz}, \citenamefont {Guloy},\ and\
		\citenamefont {Chu}}]{Gooch2009PRB}%
	\BibitemOpen
	\bibfield  {author} {\bibinfo {author} {\bibfnamefont {M.}~\bibnamefont
			{Gooch}}, \bibinfo {author} {\bibfnamefont {B.}~\bibnamefont {Lv}}, \bibinfo
		{author} {\bibfnamefont {B.}~\bibnamefont {Lorenz}}, \bibinfo {author}
		{\bibfnamefont {A.~M.}\ \bibnamefont {Guloy}}, \ and\ \bibinfo {author}
		{\bibfnamefont {C.-W.}\ \bibnamefont {Chu}},\ }\href {\doibase
		10.1103/PhysRevB.79.104504} {\bibfield  {journal} {\bibinfo  {journal} {Phys.
				Rev. B}\ }\textbf {\bibinfo {volume} {79}},\ \bibinfo {pages} {104504}
		(\bibinfo {year} {2009})}\BibitemShut {NoStop}%
	\bibitem [{\citenamefont {Gooch}\ \emph {et~al.}(2010)\citenamefont {Gooch},
		\citenamefont {Lv}, \citenamefont {Lorenz}, \citenamefont {Guloy},\ and\
		\citenamefont {Chu}}]{Gooch2010}%
	\BibitemOpen
	\bibfield  {author} {\bibinfo {author} {\bibfnamefont {M.}~\bibnamefont
			{Gooch}}, \bibinfo {author} {\bibfnamefont {B.}~\bibnamefont {Lv}}, \bibinfo
		{author} {\bibfnamefont {B.}~\bibnamefont {Lorenz}}, \bibinfo {author}
		{\bibfnamefont {A.~M.}\ \bibnamefont {Guloy}}, \ and\ \bibinfo {author}
		{\bibfnamefont {C.~W.}\ \bibnamefont {Chu}},\ }\href {\doibase
		10.1063/1.3362932} {\bibfield  {journal} {\bibinfo  {journal} {Journal of
				Applied Physics}\ }\textbf {\bibinfo {volume} {107}},\ \bibinfo {pages}
		{09E145} (\bibinfo {year} {2010})}\BibitemShut {NoStop}%
	\bibitem [{\citenamefont {Maiwald}\ \emph {et~al.}(2012)\citenamefont
		{Maiwald}, \citenamefont {Jeevan},\ and\ \citenamefont
		{Gegenwart}}]{Maiwald2012PRB}%
	\BibitemOpen
	\bibfield  {author} {\bibinfo {author} {\bibfnamefont {J.}~\bibnamefont
			{Maiwald}}, \bibinfo {author} {\bibfnamefont {H.~S.}\ \bibnamefont {Jeevan}},
		\ and\ \bibinfo {author} {\bibfnamefont {P.}~\bibnamefont {Gegenwart}},\
	}\href {\doibase 10.1103/PhysRevB.85.024511} {\bibfield  {journal} {\bibinfo
		{journal} {Phys. Rev. B}\ }\textbf {\bibinfo {volume} {85}},\ \bibinfo
	{pages} {024511} (\bibinfo {year} {2012})}\BibitemShut {NoStop}%
\bibitem [{\citenamefont {Arsenijevi\ifmmode~\acute{c}\else \'{c}\fi{}}\ \emph
	{et~al.}(2013)\citenamefont {Arsenijevi\ifmmode~\acute{c}\else \'{c}\fi{}},
	\citenamefont {Hodovanets}, \citenamefont {Ga\'al}, \citenamefont {Forr\'o},
	\citenamefont {Bud'ko},\ and\ \citenamefont {Canfield}}]{Arsenijevic2013PRB}%
\BibitemOpen
\bibfield  {author} {\bibinfo {author} {\bibfnamefont {S.}~\bibnamefont
		{Arsenijevi\ifmmode~\acute{c}\else \'{c}\fi{}}}, \bibinfo {author}
	{\bibfnamefont {H.}~\bibnamefont {Hodovanets}}, \bibinfo {author}
	{\bibfnamefont {R.}~\bibnamefont {Ga\'al}}, \bibinfo {author} {\bibfnamefont
		{L.}~\bibnamefont {Forr\'o}}, \bibinfo {author} {\bibfnamefont {S.~L.}\
		\bibnamefont {Bud'ko}}, \ and\ \bibinfo {author} {\bibfnamefont {P.~C.}\
		\bibnamefont {Canfield}},\ }\href {\doibase 10.1103/PhysRevB.87.224508}
{\bibfield  {journal} {\bibinfo  {journal} {Phys. Rev. B}\ }\textbf {\bibinfo
		{volume} {87}},\ \bibinfo {pages} {224508} (\bibinfo {year}
	{2013})}\BibitemShut {NoStop}%
\bibitem [{\citenamefont {Liu}\ \emph {et~al.}(2009{\natexlab{a}})\citenamefont
	{Liu}, \citenamefont {Kondo}, \citenamefont {Ni}, \citenamefont {Palczewski},
	\citenamefont {Bostwick}, \citenamefont {Samolyuk}, \citenamefont {Khasanov},
	\citenamefont {Shi}, \citenamefont {Rotenberg}, \citenamefont {Bud'ko},
	\citenamefont {Canfield},\ and\ \citenamefont {Kaminski}}]{Liu2009PRL}%
\BibitemOpen
\bibfield  {author} {\bibinfo {author} {\bibfnamefont {C.}~\bibnamefont
		{Liu}}, \bibinfo {author} {\bibfnamefont {T.}~\bibnamefont {Kondo}}, \bibinfo
	{author} {\bibfnamefont {N.}~\bibnamefont {Ni}}, \bibinfo {author}
	{\bibfnamefont {A.~D.}\ \bibnamefont {Palczewski}}, \bibinfo {author}
	{\bibfnamefont {A.}~\bibnamefont {Bostwick}}, \bibinfo {author}
	{\bibfnamefont {G.~D.}\ \bibnamefont {Samolyuk}}, \bibinfo {author}
	{\bibfnamefont {R.}~\bibnamefont {Khasanov}}, \bibinfo {author}
	{\bibfnamefont {M.}~\bibnamefont {Shi}}, \bibinfo {author} {\bibfnamefont
		{E.}~\bibnamefont {Rotenberg}}, \bibinfo {author} {\bibfnamefont {S.~L.}\
		\bibnamefont {Bud'ko}}, \bibinfo {author} {\bibfnamefont {P.~C.}\
		\bibnamefont {Canfield}}, \ and\ \bibinfo {author} {\bibfnamefont
		{A.}~\bibnamefont {Kaminski}},\ }\href {\doibase
	10.1103/PhysRevLett.102.167004} {\bibfield  {journal} {\bibinfo  {journal}
		{Phys. Rev. Lett.}\ }\textbf {\bibinfo {volume} {102}},\ \bibinfo {pages}
	{167004} (\bibinfo {year} {2009}{\natexlab{a}})}\BibitemShut {NoStop}%
\bibitem [{\citenamefont {Liu}\ \emph {et~al.}(2009{\natexlab{b}})\citenamefont
	{Liu}, \citenamefont {Liu}, \citenamefont {Zhao}, \citenamefont {Zhang},
	\citenamefont {Jia}, \citenamefont {Meng}, \citenamefont {Dong},
	\citenamefont {Zhang}, \citenamefont {Chen}, \citenamefont {Wang},
	\citenamefont {Zhou}, \citenamefont {Zhu}, \citenamefont {Wang},
	\citenamefont {Xu}, \citenamefont {Chen},\ and\ \citenamefont
	{Zhou}}]{Liu2009PRB}%
\BibitemOpen
\bibfield  {author} {\bibinfo {author} {\bibfnamefont {G.}~\bibnamefont
		{Liu}}, \bibinfo {author} {\bibfnamefont {H.}~\bibnamefont {Liu}}, \bibinfo
	{author} {\bibfnamefont {L.}~\bibnamefont {Zhao}}, \bibinfo {author}
	{\bibfnamefont {W.}~\bibnamefont {Zhang}}, \bibinfo {author} {\bibfnamefont
		{X.}~\bibnamefont {Jia}}, \bibinfo {author} {\bibfnamefont {J.}~\bibnamefont
		{Meng}}, \bibinfo {author} {\bibfnamefont {X.}~\bibnamefont {Dong}}, \bibinfo
	{author} {\bibfnamefont {J.}~\bibnamefont {Zhang}}, \bibinfo {author}
	{\bibfnamefont {G.~F.}\ \bibnamefont {Chen}}, \bibinfo {author}
	{\bibfnamefont {G.}~\bibnamefont {Wang}}, \bibinfo {author} {\bibfnamefont
		{Y.}~\bibnamefont {Zhou}}, \bibinfo {author} {\bibfnamefont {Y.}~\bibnamefont
		{Zhu}}, \bibinfo {author} {\bibfnamefont {X.}~\bibnamefont {Wang}}, \bibinfo
	{author} {\bibfnamefont {Z.}~\bibnamefont {Xu}}, \bibinfo {author}
	{\bibfnamefont {C.}~\bibnamefont {Chen}}, \ and\ \bibinfo {author}
	{\bibfnamefont {X.~J.}\ \bibnamefont {Zhou}},\ }\href {\doibase
	10.1103/PhysRevB.80.134519} {\bibfield  {journal} {\bibinfo  {journal} {Phys.
			Rev. B}\ }\textbf {\bibinfo {volume} {80}},\ \bibinfo {pages} {134519}
	(\bibinfo {year} {2009}{\natexlab{b}})}\BibitemShut {NoStop}%
\bibitem [{\citenamefont {Dhaka}\ \emph {et~al.}(2013)\citenamefont {Dhaka},
	\citenamefont {Hahn}, \citenamefont {Razzoli}, \citenamefont {Jiang},
	\citenamefont {Shi}, \citenamefont {Harmon}, \citenamefont {Thaler},
	\citenamefont {Bud'ko}, \citenamefont {Canfield},\ and\ \citenamefont
	{Kaminski}}]{Dhaka2013PRL}%
\BibitemOpen
\bibfield  {author} {\bibinfo {author} {\bibfnamefont {R.~S.}\ \bibnamefont
		{Dhaka}}, \bibinfo {author} {\bibfnamefont {S.~E.}\ \bibnamefont {Hahn}},
	\bibinfo {author} {\bibfnamefont {E.}~\bibnamefont {Razzoli}}, \bibinfo
	{author} {\bibfnamefont {R.}~\bibnamefont {Jiang}}, \bibinfo {author}
	{\bibfnamefont {M.}~\bibnamefont {Shi}}, \bibinfo {author} {\bibfnamefont
		{B.~N.}\ \bibnamefont {Harmon}}, \bibinfo {author} {\bibfnamefont
		{A.}~\bibnamefont {Thaler}}, \bibinfo {author} {\bibfnamefont {S.~L.}\
		\bibnamefont {Bud'ko}}, \bibinfo {author} {\bibfnamefont {P.~C.}\
		\bibnamefont {Canfield}}, \ and\ \bibinfo {author} {\bibfnamefont
		{A.}~\bibnamefont {Kaminski}},\ }\href {\doibase
	10.1103/PhysRevLett.110.067002} {\bibfield  {journal} {\bibinfo  {journal}
		{Phys. Rev. Lett.}\ }\textbf {\bibinfo {volume} {110}},\ \bibinfo {pages}
	{067002} (\bibinfo {year} {2013})}\BibitemShut {NoStop}%
\bibitem [{\citenamefont {Hassinger}\ \emph {et~al.}(2012)\citenamefont
	{Hassinger}, \citenamefont {Gredat}, \citenamefont {Valade}, \citenamefont
	{de~Cotret}, \citenamefont {Juneau-Fecteau}, \citenamefont {Reid},
	\citenamefont {Kim}, \citenamefont {Tanatar}, \citenamefont {Prozorov},
	\citenamefont {Shen}, \citenamefont {Wen}, \citenamefont {Doiron-Leyraud},\
	and\ \citenamefont {Taillefer}}]{Hassinger2012PRB}%
\BibitemOpen
\bibfield  {author} {\bibinfo {author} {\bibfnamefont {E.}~\bibnamefont
		{Hassinger}}, \bibinfo {author} {\bibfnamefont {G.}~\bibnamefont {Gredat}},
	\bibinfo {author} {\bibfnamefont {F.}~\bibnamefont {Valade}}, \bibinfo
	{author} {\bibfnamefont {S.~R.}\ \bibnamefont {de~Cotret}}, \bibinfo {author}
	{\bibfnamefont {A.}~\bibnamefont {Juneau-Fecteau}}, \bibinfo {author}
	{\bibfnamefont {J.-P.}\ \bibnamefont {Reid}}, \bibinfo {author}
	{\bibfnamefont {H.}~\bibnamefont {Kim}}, \bibinfo {author} {\bibfnamefont
		{M.~A.}\ \bibnamefont {Tanatar}}, \bibinfo {author} {\bibfnamefont
		{R.}~\bibnamefont {Prozorov}}, \bibinfo {author} {\bibfnamefont
		{B.}~\bibnamefont {Shen}}, \bibinfo {author} {\bibfnamefont {H.-H.}\
		\bibnamefont {Wen}}, \bibinfo {author} {\bibfnamefont {N.}~\bibnamefont
		{Doiron-Leyraud}}, \ and\ \bibinfo {author} {\bibfnamefont {L.}~\bibnamefont
		{Taillefer}},\ }\href {\doibase 10.1103/PhysRevB.86.140502} {\bibfield
	{journal} {\bibinfo  {journal} {Phys. Rev. B}\ }\textbf {\bibinfo {volume}
		{86}},\ \bibinfo {pages} {140502} (\bibinfo {year} {2012})}\BibitemShut
{NoStop}%
\bibitem [{\citenamefont {Hassinger}\ \emph {et~al.}(2016)\citenamefont
	{Hassinger}, \citenamefont {Gredat}, \citenamefont {Valade}, \citenamefont
	{de~Cotret}, \citenamefont {Cyr-Choini\`ere}, \citenamefont {Juneau-Fecteau},
	\citenamefont {Reid}, \citenamefont {Kim}, \citenamefont {Tanatar},
	\citenamefont {Prozorov}, \citenamefont {Shen}, \citenamefont {Wen},
	\citenamefont {Doiron-Leyraud},\ and\ \citenamefont
	{Taillefer}}]{Hassinger2016PRB}%
\BibitemOpen
\bibfield  {author} {\bibinfo {author} {\bibfnamefont {E.}~\bibnamefont
		{Hassinger}}, \bibinfo {author} {\bibfnamefont {G.}~\bibnamefont {Gredat}},
	\bibinfo {author} {\bibfnamefont {F.}~\bibnamefont {Valade}}, \bibinfo
	{author} {\bibfnamefont {S.~R.}\ \bibnamefont {de~Cotret}}, \bibinfo {author}
	{\bibfnamefont {O.}~\bibnamefont {Cyr-Choini\`ere}}, \bibinfo {author}
	{\bibfnamefont {A.}~\bibnamefont {Juneau-Fecteau}}, \bibinfo {author}
	{\bibfnamefont {J.-P.}\ \bibnamefont {Reid}}, \bibinfo {author}
	{\bibfnamefont {H.}~\bibnamefont {Kim}}, \bibinfo {author} {\bibfnamefont
		{M.~A.}\ \bibnamefont {Tanatar}}, \bibinfo {author} {\bibfnamefont
		{R.}~\bibnamefont {Prozorov}}, \bibinfo {author} {\bibfnamefont
		{B.}~\bibnamefont {Shen}}, \bibinfo {author} {\bibfnamefont {H.-H.}\
		\bibnamefont {Wen}}, \bibinfo {author} {\bibfnamefont {N.}~\bibnamefont
		{Doiron-Leyraud}}, \ and\ \bibinfo {author} {\bibfnamefont {L.}~\bibnamefont
		{Taillefer}},\ }\href {\doibase 10.1103/PhysRevB.93.144401} {\bibfield
	{journal} {\bibinfo  {journal} {Phys. Rev. B}\ }\textbf {\bibinfo {volume}
		{93}},\ \bibinfo {pages} {144401} (\bibinfo {year} {2016})}\BibitemShut
{NoStop}%
\bibitem [{\citenamefont {Thompson}(1984)}]{Thompson1984}%
\BibitemOpen
\bibfield  {author} {\bibinfo {author} {\bibfnamefont {J.~D.}\ \bibnamefont
		{Thompson}},\ }\href {http://aip.scitation.org/doi/10.1063/1.1137730}
{\bibfield  {journal} {\bibinfo  {journal} {Review of Scientific
			Instruments}\ } (\bibinfo {year} {1984})}\BibitemShut {NoStop}%
\end{thebibliography}
%

\end{document}